\documentclass[12pt,preprint]{aastex}

\slugcomment{AJ, Vol.132, p.781-800, 2006 August}

\shorttitle{2MASS Near-Infrared Colors of Magellanic Cloud Star Clusters}
\shortauthors{Pessev et al.}

\begin{document}

\title{A Database of 2MASS Near-Infrared Colors of 
Magellanic Cloud Star Clusters}
\author{Peter M. Pessev, Paul Goudfrooij, Thomas H. Puzia, \& Rupali
  Chandar\altaffilmark{1}} 
\affil{Space Telescope Science Institute}
\affil{3700 San Martin Drive, Baltimore, MD 21218, USA}
\email{pessev, goudfroo, tpuzia, rupali@stsci.edu}

\altaffiltext{1}{Present Address: Department of Physics and Astronomy, Johns
  Hopkins University, 3400 N. Charles Street, Baltimore, MD 21218} 

\begin{abstract}
The (rest-frame) near-IR domain contains important stellar population
diagnostics and is often used to estimate masses of galaxies at low as
well as high redshifts. However, many stellar population models are still
relatively poorly calibrated in this part of the spectrum. To allow an
improvement of this calibration we present a new database of integrated
near-infrared {\it JHK$_{s}$} magnitudes for 75 star clusters in the
Magellanic Clouds, using the 2-Micron All-Sky Survey (2MASS). The majority
of the clusters in our sample have robust age and metallicity estimates
from color-magnitude diagrams available in the literature, and populate a
range of ages from 10 Myr to 15 Gyr and a range in [Fe/H] from $-2.17$ to
$+0.01$ dex. A comparison with matched star clusters in the 2MASS Extended
Source Catalog (XSC) reveals that the XSC only provides a good fit to the
{\it unresolved} component of the cluster stellar population. We also
compare our results with the often-cited single-channel {\it JHK}
photometry of \cite{persson83}, and find significant differences,
especially for their 30\arcsec-diameter apertures up to $\sim\!2.5$ mag in
the $K$-band, more than 1 mag in $J\!-\!K$, and up to 0.5 mag in
$H\!-\!K$. Using simulations to center apertures based on maximum light
throughput (as performed by \citeauthor{persson83}), we show that these
differences can be attributed to near-IR-bright cluster stars (e.g.,
Carbon stars) located away from the true center of the star clusters. The
wide age and metallicity coverage of our integrated {\it JHK$_s$} photometry
sample constitutes a fundamental dataset for testing population synthesis
model predictions, and for direct comparison with near-IR observations of
distant stellar populations.
\end{abstract}

%%%%%%%%%%%%%%%%%%%%%%%%%%%%%%%%%%%%%
\section{Introduction} 
Much of our understanding of galaxy formation and evolution comes from
studying stellar populations in different galaxy types, both in the
present and early universe. Two key parameters of stellar systems which
are widely used throughout the literature are mean ages and metallicities.
Ages and/or metallicities of stellar systems in photometric surveys are
estimated by comparing measured integrated colors with the predictions of
evolutionary synthesis models \citep[e.g.][]{bc93, bc03, worthey94,
vazdekis99, maraston98, maraston05}. These models utilize stellar
isochrone libraries, which are synthesized in appropriate combinations to
represent stellar systems at different ages and metallicities. There are
however, two important limitations inherent to these models. First, the
stellar libraries themselves contain mostly stars in the solar
neighborhood, which have a star formation history that is not necessarily
typical for extragalactic populations (e.g., relatively little variation
in chemical composition). Second, the synthesis models oversimplify the
more rapid (but very luminous) phases of stellar evolution (e.g.,
thermally pulsing asymptotic giant branch stars). Given the very
fundamental nature of the information that is derived by comparison with
these models, it is imperative that population synthesis models be as
accurate as possible.

Simple stellar population (SSP) models are empirically calibrated to
observations of real star clusters for which ages and metallicities are
known from independent analysis, e.g., color-magnitude diagrams
\citep[e.g.][]{bruzual97, maraston03}. While much of the work to date has
been carried out at optical wavelengths, the near-infrared (NIR) regime
contains some very important diagnostics for deriving basic properties of
stellar systems. In fact, this wavelength regime has been shown to be very
important for sorting out the effects of age and metallicity, particularly
in stellar populations older than about 300 Myr
\citep[e.g.][]{goudfrooij01, puzia02, hempel04}. Due to recent advances in
the instrumentation and detector capabilities in the NIR passbands, and
considering the focus on the infrared in the next generation of
telescopes, it is clear that the accuracy of SSP models in the NIR is
going to be even more important in the future.

In this work, we present integrated NIR colors of a large sample of
star clusters in the Large and Small Magellanic Cloud (hereafter LMC
and SMC). We make use of data from the Two Micron All Sky Survey
\citep[2MASS][]{skrutskie97}, which offers uniform, high-quality imaging of the entire sky
in three bands, $J$ (1.25 $\mu$m), $H$ (1.65 $\mu$m), and
$K_s$\footnote{For a description of the "$K$ short" ($K_s$) band, see
  \cite{persson98}.} (2.16 $\mu$m). Our main goal is to provide a new
database of intrinsic NIR magnitudes and colors of clusters with
well-known ages and metallicities from deep color-magnitude diagrams
(CMDs), that can be utilized as a calibration dataset by existing and
future generation SSP models. The clusters in the Magellanic Clouds
are very suitable for addressing this issue. They cover a wide range
of ages, and they are close enough for detailed CMD studies using the
Hubble Space Telescope {\it (HST)}, (in some cases also with large
telescopes from the ground). Unlike the globular cluster system (GCS)
of our Galaxy, there are a significant number of objects with
intermediate ages ($0.3 - 3$ Gyr) in the LMC and SMC. The
integrated-light properties of these systems are affected strongly by
AGB stars which are extremely luminous in the NIR, and their
contribution to the light in that part of the spectrum is largely
underestimated by most existing SSP models \citep[see][]{maraston05}. 

The measurement of integrated magnitudes and colors of star clusters in
the Magellanic Clouds is complicated by several factors. One problem is
that of accurate centering of the aperture. Many of these
clusters are superposed onto a relatively high surface density of stars
associated with the LMC or SMC, and some have a rather irregular field
distribution and/or are not particularly symmetric due to the
superposition of bright stars (be it supergiants or AGB stars, associated
with the cluster itself, those from the body of the LMC or SMC, or
Galactic foreground stars). On the other hand, it should be recognized
that the use of 2-dimensional imagery renders these problems much less
severe than they were for often-cited previous studies which used single-channel
photometers and diaphragms which were centered either by eye or by maximum
throughput.

The present study is complementary to the information about Magellanic Cloud clusters in 2MASS Extended Source Catalog \citep[XSC][]{jarrett2000} in three ways: {\it (i)} providing photometry for a set of clusters that are not present in the 2MASS XSC; {\it (ii)} we take into account the flux from the point sources associated with the star clusters, which are rejected by the XSC pipeline (see \S3.2 for details); {\it (iii)} better sampling of the curves of growth with a step of 1\arcsec , instead of 11 fixed circular apertures.

This paper is organized as follows: Section 2 describes the sample
selection, data acquisition and reduction. The results, including
comparison with previous works and  2MASS XSC are presented in \S3.
Finally, a summary is provided in \S4.

%%%%%%%%%%%%%%%%%%%%%%%%%%%%%%%%%%%%%
\section{Near-Infrared Data}
\subsection{Sample selection}
Our original sample of star clusters was adopted from \cite{mackey03a,
  mackey03b}, and most have accurate CMD ages and metallicities from
the literature. We particularly pay attention to the largest possible
coverage of the available age/metallicity parameter space. In
addition, we select intermediate-age and young clusters which have no
known counterparts in the Milky Way globular cluster system. The
adopted distance moduli are $m-M = 18.89$ and $m-M = 18.50$ for the
SMC \citep{harries2003} and LMC \citep{alves2004} respectively. Basic information for all
objects is provided in Tables~\ref{tab:smc} and \ref{tab:lmc} (respectively for
star clusters in the SMC and LMC). 

The young SMC cluster NGC~176 was included in the original list, but
after inspection of the 2MASS images it became clear that the NIR data
is too shallow to derive reliable integrated colors. R136 in LMC, the
youngest object in the preliminary selection, is embedded in an
extensive emission region that would affect the results of the
integrated photometry. We decided not to include these two clusters in
the final list.  $J$, $H$, and $K_s$ postage stamp images of
representative objects in our sample are presented in
Figure~\ref{fig:smc} (SMC) and \ref{fig:lmc} (LMC). $V$-band frames
for the majority of the SMC objects, included in this work, can be
found in \cite{hill05}. 

%%%%%%%%%%%%%%%%%%%%%%%%%%%%%%%%%%%%%
\subsection{2MASS Atlas Images}
The 2MASS Atlas Images originate from 6$^\circ$-long survey scans 
using an effective integration time of 7.8 seconds per tile. 
$J$, $H$, and $K_s$ images were retrieved using the 2MASS
interactive image
service\footnote{http://irsa.ipac.caltech.edu/applications/2MASS/IM/intera
  ctive.html}. The queries were usually sent by object name and in
some cases, when the name qualifier was not recognized, by
coordinates. In most cases an object could be found on several sets of
frames, allowing us to choose the best one, taking into account the
relative position of the cluster and the characteristics of each
field. Tables~\ref{tab:smcAtl} and \ref{tab:lmcAtl} provide
information on the Atlas Images, selected for our study, for SMC and
LMC clusters, respectively. Column~2 in these tables provides the
number of different sets of images retrieved for each object. 

The raw survey data was reduced at the Infrared Processing and
Analysis Center with the pipeline specifically developed for
2MASS. The imaging data is resampled to 1\arcsec / pixel, calibrated to one second 
integration time and contain both the
astrometric solution and the photometric zero points for each
individual Atlas Image \citep{cutri03}. The astrometric solutions
are obtained in the International Celestial Reference System via the
Tycho-2 Reference Catalog. Taking into account the higher value of the extended
source uncertainty \citep{cutri03}, all cluster positions derived in the present work
were rounded to the nearest half pixel (0\farcs5). 

The photometric zero points are based on observations of fields,
covering the NIR standards from the list of \cite{persson98} or the
UKIRT group of faint, equatorial NIR standard stars
\citep{casali92}. The solution is
derived independently for each band and minimizes the residuals by a
least square fit of the zero point, airmass, and atmospheric
extinction \citep{nikolaev2000}.  The distributions of the zero point differences for all
standard fields in all survey nights turned out to be Gaussian with
RMS residual values of 0.011, 0.007 and 0.007 magnitudes in $J$,$H$
and $K_s$, respectively \citep{cutri03}. These values are added in
quadrature to the photometric errors in this work.

%%%%%%%%%%%%%%%%%%%%%%%%%%%%%%%%%%%%%
\subsection{Data analysis}
In order to measure the integrated cluster magnitudes, the following multistep procedure was applied to each object of the sample:{\it(i)} PSF-fitting photometry of the point sources; {\it(ii)} determination of the center position for the integrated curve of growth photometry; {\it(iii)} subtraction of the background/foreground point source luminosity function (LF) from the LF of the cluster field; {\it(iv)} integrated photometry of the total, background-subtracted and unresolved component of the object in each survey band; {\it(v)} calculation of the photometric errors for each measurement aperture.

For basic data analysis we use the suite of IRAF\footnote{IRAF is
distributed by the National Optical Astronomical Observatory, which is
operated by the Associations of Universities for Research in Astronomy
Inc., under cooperative agreement with the National Science Foundation.}
tasks and perform DAOPHOTII/ALLSTAR photometry \citep{stetson87} on each
frame. Typically several bright and well-isolated stars were used to
construct the PSF for each frame. As this is only an intermediate stage in
the process of deriving the total integrated cluster magnitudes, aperture
corrections are not applied at this point. The ALLSTAR routine is used to
produce frames on which individual stars are removed after being measured.
We will refer to these frames as "residual frames", which will later be
used to study the part of the cluster stellar population not resolved in
the 2MASS images.

\begin{figure}
\centering
\includegraphics[bb=14 14 535 706,width=14cm]{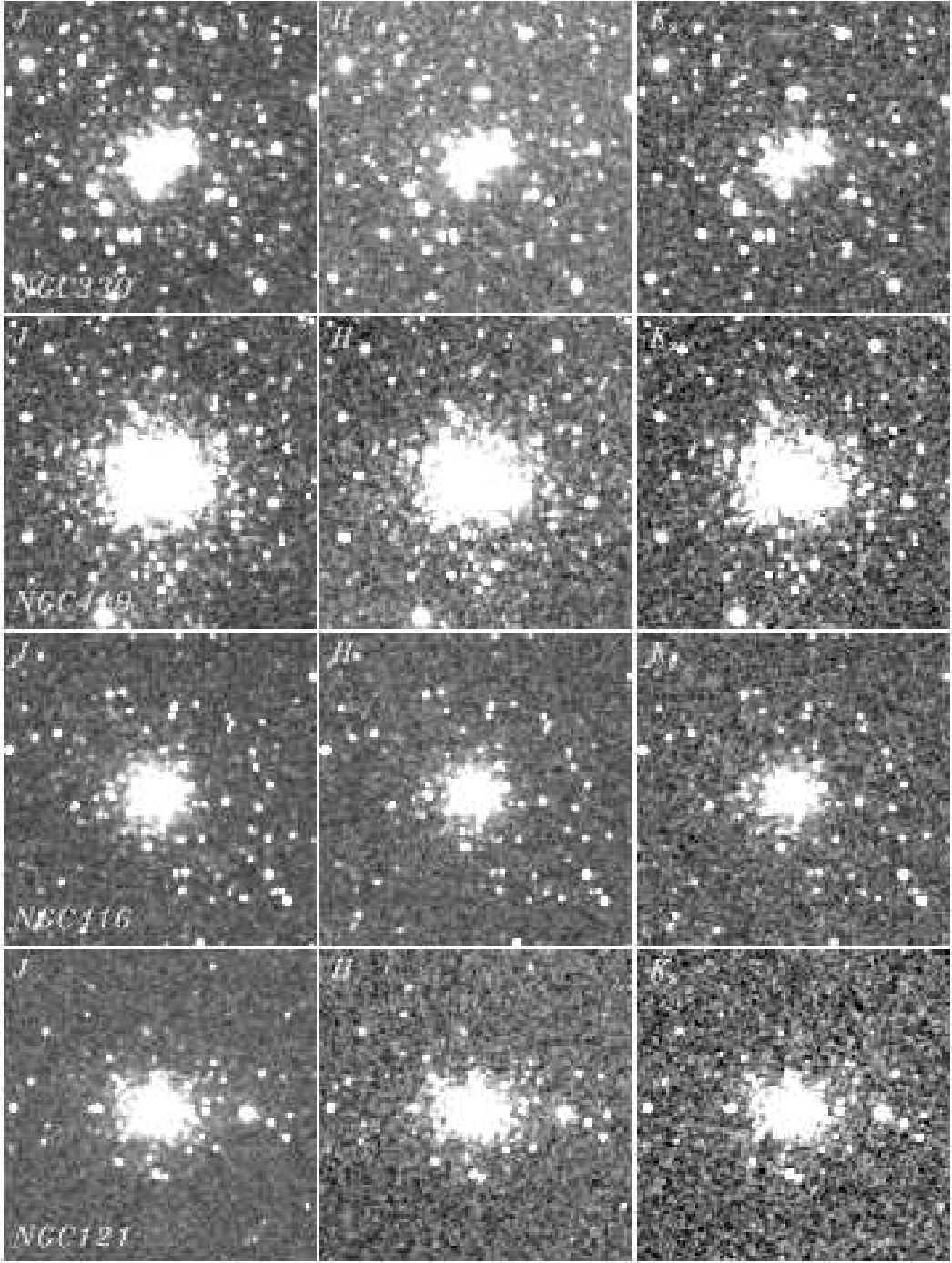}
\caption{$J$,$H$, and $K_s$ images of 4 SMC clusters. Each image is 200\arcsec\
  $\times$ 200\arcsec, centered on the cluster position derived in the present
  paper. North is up, East is to the left. The curves of growth for each of
  these objects can be found on Figure~\ref{fig:cg4smc}.}
\label{fig:smc}
\end{figure}

\begin{figure}
\centering
\includegraphics[bb=14 14 535 706,width=14cm]{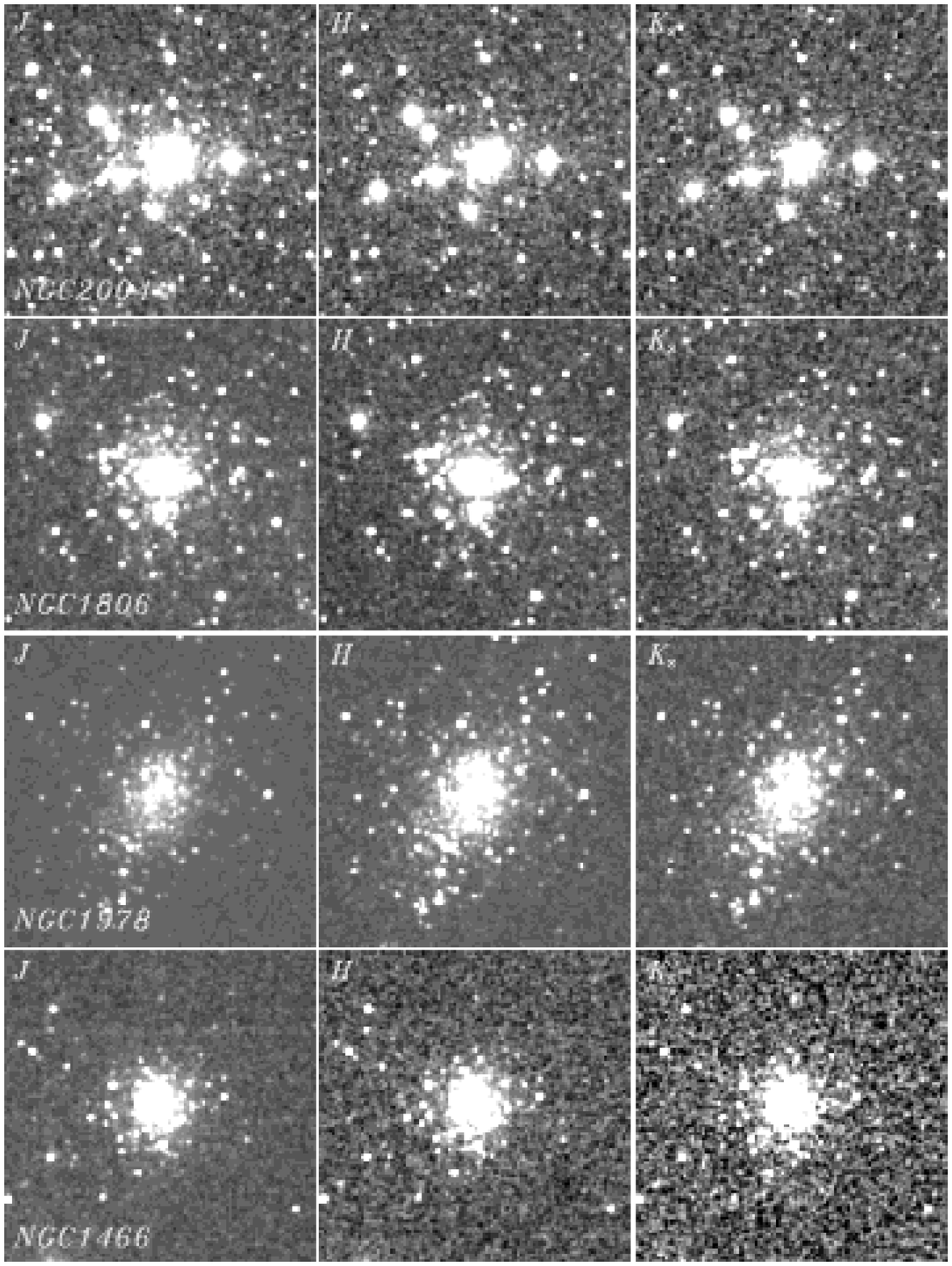}
\caption{$J$,$H$, and $K_s$ images of 4 LMC clusters. Each image is 200\arcsec\
  $\times$ 200\arcsec, centered on the cluster position derived in the present
  paper. North is up, East is to the left. The curves of growth for each of
  these objects can be found on Figure~\ref{fig:cg4lmc}.}
\label{fig:lmc}
\end{figure}

In many cases the coordinates taken from cluster catalogs in the
literature do not provide an accurate position for the center of
individual clusters. We applied a simple and robust method to derive
the centers in the present paper. The original frames and the residual
frames in each survey band were smoothed with a large Gaussian
kernel. The size of the kernel varied as a function of the cluster
size on the $J$ frames, where the sensitivity of 2MASS reaches its
peak. The maximum flux values on an image subsection of the smoothed
images in each of the survey bands, derived by the IRAF task MINMAX
were used to determine the individual cluster centers in $J,H,$ and
$K_s$. They were later averaged to derive the final center coordinates
that were used for the integrated photometry in the present
paper. Possible sources of confusion (e.g., bright stars outside the
cluster area, present on the original Atlas frames) were avoided by
performing the procedure described earlier on an image subsection,
covering the rough cluster position on the smoothed images. We point
out that the cluster images in Figures~\ref{fig:smc} and \ref{fig:lmc}
are extracted from the 2MASS frames, using the centers derived with
this procedure. Each image covers 200\arcsec\ $\times$ 200\arcsec\ and
is representative of the size of the largest aperture used to measure
integrated cluster magnitudes. 

In a few cases the object was situated close to the edge of the Atlas
frame. This is not a serious problem for most clusters because we were
still able to sample the flux out to large enough radii to derive the
total integrated magnitudes. One exception was Lindsay~1 which was
split almost equally between two sets of frames. The special data
reduction procedure, applied to this object is described in
Section~\ref{notes} below. 

%%%%%%%%%%%%%%%%%%%%%%%%%%%%%%%%%%%%%
\subsection{Stellar Background Subtraction}
\begin{figure}
\centering
\includegraphics[bb=0 85 566 651,width=12cm]{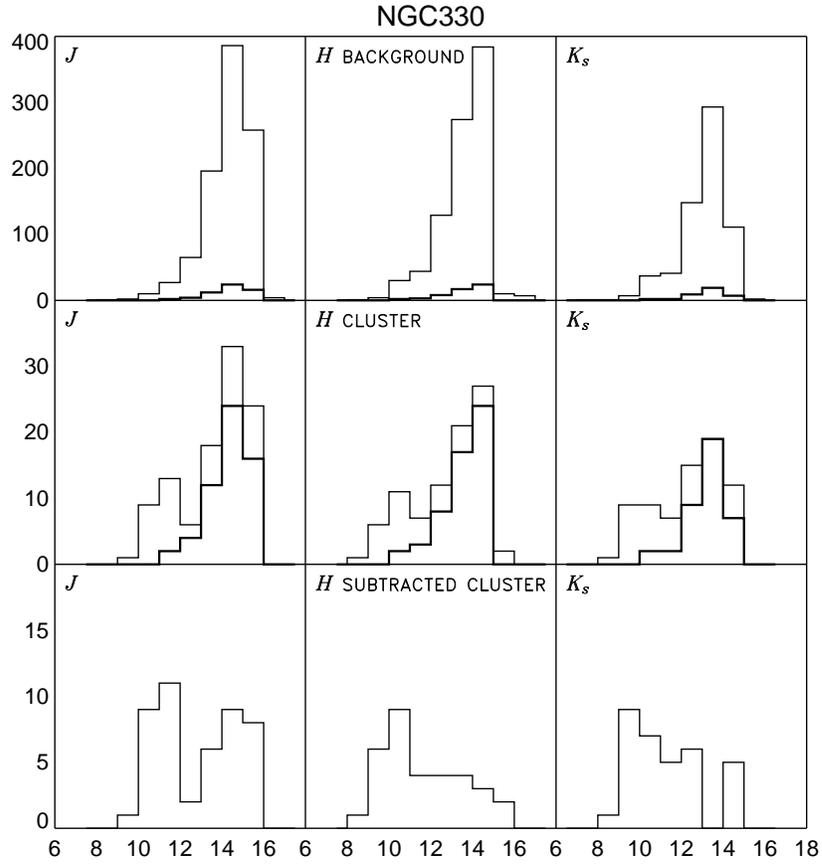}
\caption{Luminosity functions of the background, cluster field and the cluster
  field after background subtraction from top to bottom. The thick line on the
  first and second row of pannels is representing the background LF scaled to
  the area of the largest aperture.}
\label{fig:bkg}
\end{figure}

To estimate the contamination of our globular cluster fields by
foreground stars and stars associated with the body of the LMC or SMC
we used the portion of the frames outside of the largest aperture used
for integrated-light photometry, (typically 100 \arcsec). The
luminosity function (LF) for the point sources in these background
regions was scaled to the cluster area used for photometry. This was
statistically subtracted from the point source LF measured inside the
photometric radius. The area of the background regions slightly varied as a function of the largest aperture size, but even in the case of 100 \arcsec aperture radii it was  more than 15 times larger than the photometry area. In this way we achieve a good estimate of the
background/foreground contamination, one which is much less affected
by local stellar variations and therefore superior to just subtracting
a normalized background flux from a neighboring annulus. To illustrate
our procedure, in Figure~\ref{fig:bkg} we show the LFs of the
background, cluster, and the cleaned LF after the background
subtraction, for the heavily contaminated cluster NGC~330. 

In cases where bright stars (not sampled by the field star LF) are present in the cluster aperture, after the background subtraction  we are utilizing their ALLSTAR PSF magnitudes and corresponding colors to make a rough estimation of the spectral and luminosity class using the
work of \cite{ducati01}. The visual magnitudes of these objects were
recovered using the visual -- infrared colors from the same study for
a certain spectral and luminosity class. The results were compared
with the age of the cluster and the expected absolute visual
magnitudes of these stars at the distance to SMC or LMC. If there was
a discrepancy between the measured and expected magnitudes or if such
stars were unlikely to be found in a cluster with the given age, they
were subtracted. There are several examples of this procedure
described in Section~\ref{notes}. 

%%%%%%%%%%%%%%%%%%%%%%%%%%%%%%%%%%%%%
\subsection{Integrated Cluster Photometry}
The aperture photometry of the clusters was performed with IRAF APPHOT package on a set of 
three residual images in each survey band. The images are a result by the application of 
DAOPHOT/ALLSTAR and the IRAF SUBSTAR procedure on the original atlas frames. On the first 
of them all detected point sources were removed (actually this is the output residual 
image from ALLSTAR), on the second the LF of the background field was subtracted from the 
cluster area using SUBSTAR with a exclude file, containing the stars remaining after the 
statistical subtraction, the third image contained all the point sources within the 
photometry aperture. The images were used to measure the flux from the unresolved 
component, background subtracted and the total flux from the object,
 using a set of apertures ranging from 1\arcsec\ to 100
\arcsec\ in radius with a step size of 1\arcsec. We computed curves of growth
for all sample globular clusters in the three survey bands. The sky
background level in each frame was estimated in a sky annulus
encircling our largest aperture with width at least 10 pixels. The exclusion of the
stars outside of the largest aperture prior to the final integrated photometry is 
providing a better estimate of the sky background. In those cases where the cluster was
situated close to the frame edge, we used the largest aperture
possible, and the background levels were measured in a nearby region
that matched or exceeded the equivalent area of a full background
annulus circle. 

For each aperture the error introduced by the stochastic fluctuations
in the stellar population of the foreground/background was
estimated. We computed the luminosity function (LF) of objects outside
each photometry aperture for each particular object. The corresponding
flux was integrated over the entire LF and the standard deviation of
stellar counts were calculated in bins of $\Delta m=1$ mag. These
values were then normalized to the area used for the cluster
photometry. Bright stars close to saturation were identified on the images and excluded
from the photometry prior to the procedures described above. The error values listed in 
Tables~\ref{tab:photsmc} and \ref{tab:photsmc} in each survey band are the quadrature 
sum of the photometry errors from APPHOT, the 2MASS zero point errors and the calculated  
background stochastic fluctuations. In
general, the errors of our photometry increase as a function of
aperture radius and depend on background properties.  

\begin{figure}
\centering
\includegraphics[bb=54 14 564 779,width=10cm]{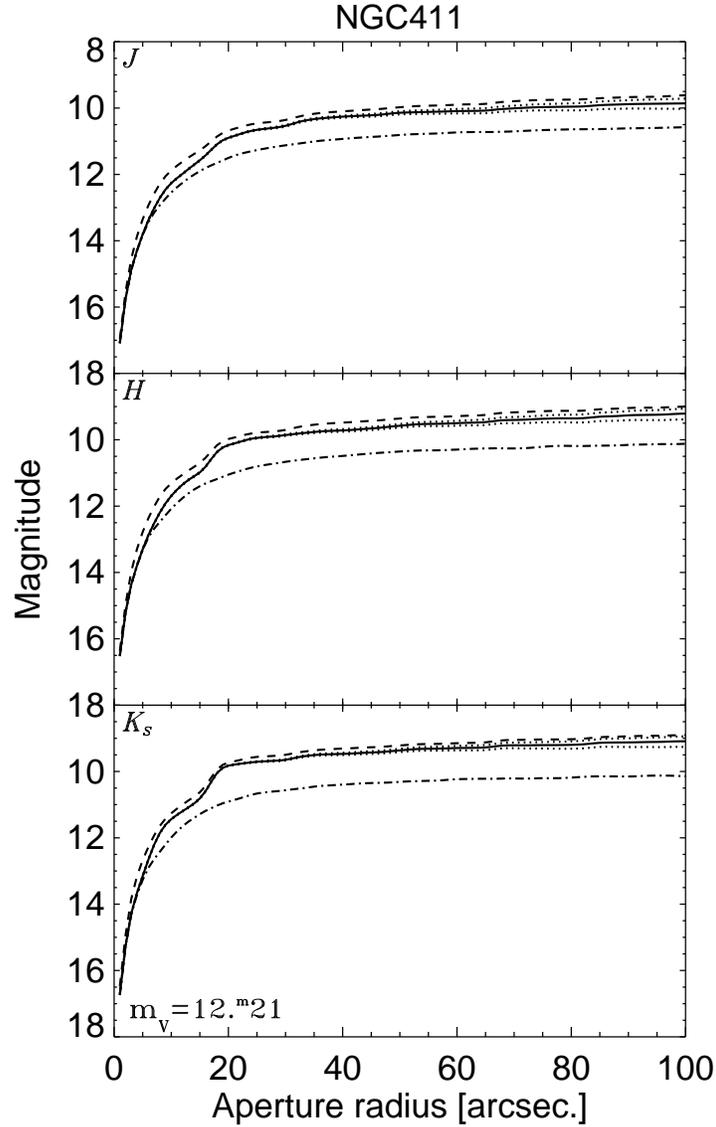}
\caption{Curves of growth in the three 2MASS bands for the SMC cluster
  NGC411. The dashed curve represents total flux from the object (no
  background/foreground subtraction applied), dots and dashes are standing for
  the unresolved component on the Atlas Images. The solid lines are showing the background subtracted curve of growth, and the errors due to the stochastic fluctuations of the background are overplotted with dotted lines. The $V$ photoelectric magnitude in a 62\arcsec diaphragm is taken from \cite{alcaino78}.}
\label{fig:n411}
\end{figure}

\begin{figure}
\centering
\includegraphics[bb=54 14 564 779,width=10cm]{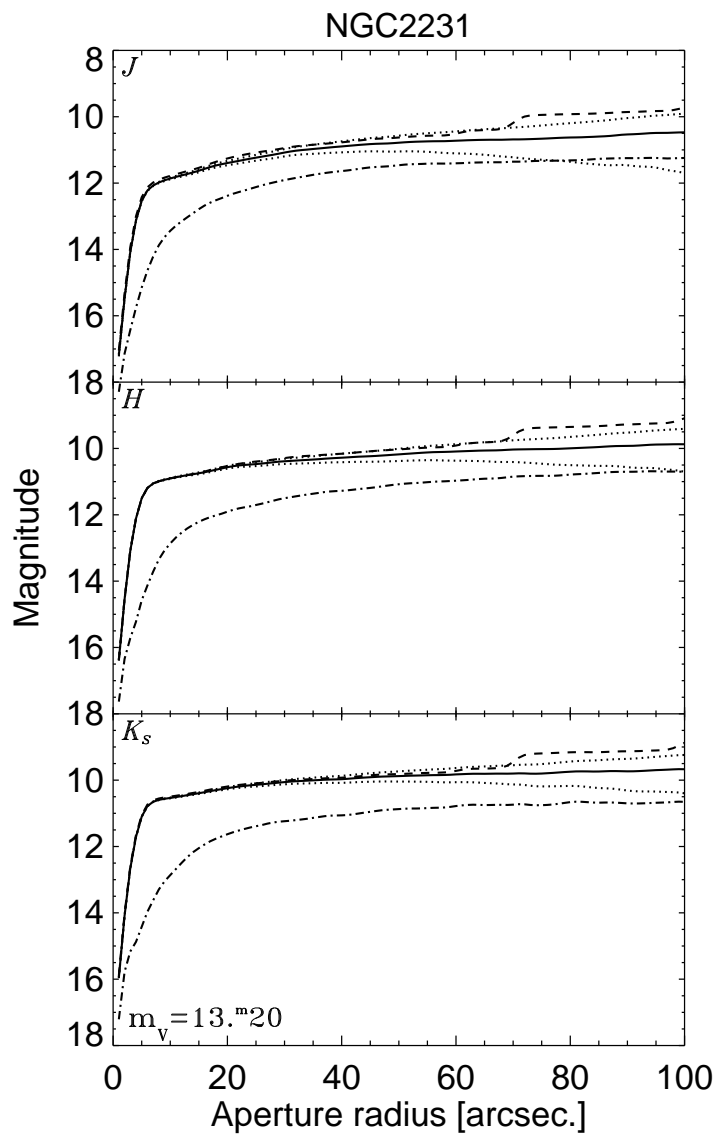}
\caption{Curves of growth in the three 2MASS bands for the LMC cluster
  NGC2231. The dashed curve represents total flux from the object (no
  background/foreground subtraction applied), note the clear ``bump'' due to a
  foreground star. Solid line is standing for the background subtracted curve of growth, dots  and dashes for the unresolved component on the Atlas Images. The errors due to the stochastic fluctuations  
  in the background are overplotted with dotted lines.The values of these errors are quite high because the bright stars in the background field are not excluded from the LF and this is illustrating the possibility of underestimation of the cluster total magnitude. The photoelectric visual magnitude in 44\arcsec aperture is taken from \cite{vdBergh81}. }
\label{fig:n2231}
\end{figure}

Example curves of
growth of the photometry for the SMC cluster NGC\,411 and LMC cluster
NGC\,2231 are presented in Figures~\ref{fig:n411} and \ref{fig:n2231},
respectively. The solid line shows the magnitude of the cluster after
background subtraction. The estimated errors due to stochastic
background fluctuations are overplotted with dotted lines. The dashed
curve represents the total flux from the cluster (without background
subtraction), and the unresolved component is plotted with dot-dashed
lines. NGC~2231 in Figure~\ref{fig:n2231} also illustrates what could be the effect on the 
integrated magnitudes of the cluster if there is improper handling of the background 
subtraction. This cluster lie in a region of rellativelly high apparent stellar density 
and there are several bright stars present. The stochastic fluctustions of the stellar 
background show the possibility of a severe overestimation of the cluster total magnitude 
if the influence of the bright stellar objects is not taken into account. Note that the 
error values  in Table~\ref{tab:photsmc} are calculated with the bright stars excluded 
from the background LF. Integrated magnitudes and photometric
curves of growth for all star clusters analyzed in this study are
available upon request from the first author.

%%%%%%%%%%%%%%%%%%%%%%%%%%%%%%%%%%%%%
\subsection{Extinction correction}
\label{ext}
In order to determine the intrinsic magnitudes and colors of our sample
clusters, the measurements must be corrected for the effect of extinction.
Cluster extinction values were obtained from three independent studies:
\cite{burstein82}, \cite{schlegel98}, and the Magellanic Clouds
Photometric Survey \citep[MCPS,][]{zaritsky97}: see \cite{zaritsky02} for
SMC and \cite{zaritsky04} for LMC. The study by \citeauthor{burstein82} is
based on maps of the HI emission, while the \citeauthor{schlegel98} maps
use IRAS/DIRBE data of the FIR sky emission. The corrections provided by
\cite{schlegel98} are superior to those of \cite{burstein82}, because of
the improved spatial resolution and the fact that they estimate the
extinction from the dust properties directly, not using HI as an
intermediate agent. But there is an important caveat - the
\citeauthor{schlegel98} maps are highly uncertain in the inner regions of
the Clouds because their temperature structure wasn't sufficiently
resolved by DIRBE. The most recent development in Magellanic Clouds
extinction work is the MCPS data. This survey covers the central
4$^\circ \times$4$^\circ$ of the SMC and 8$^\circ \times$8$^\circ$ of the LMC
in {\it UBVI}. The 
limiting magnitude of the MCPS maps (set primarily by crowding) is $V=21$
mag. However, these maps cover only the inner body of the Magellanic
Clouds and extinction information for star clusters located in the outer
regions is not available.

\begin{figure}
\centering
\includegraphics[bb= 14 14 520 700,width=12cm,angle=90]{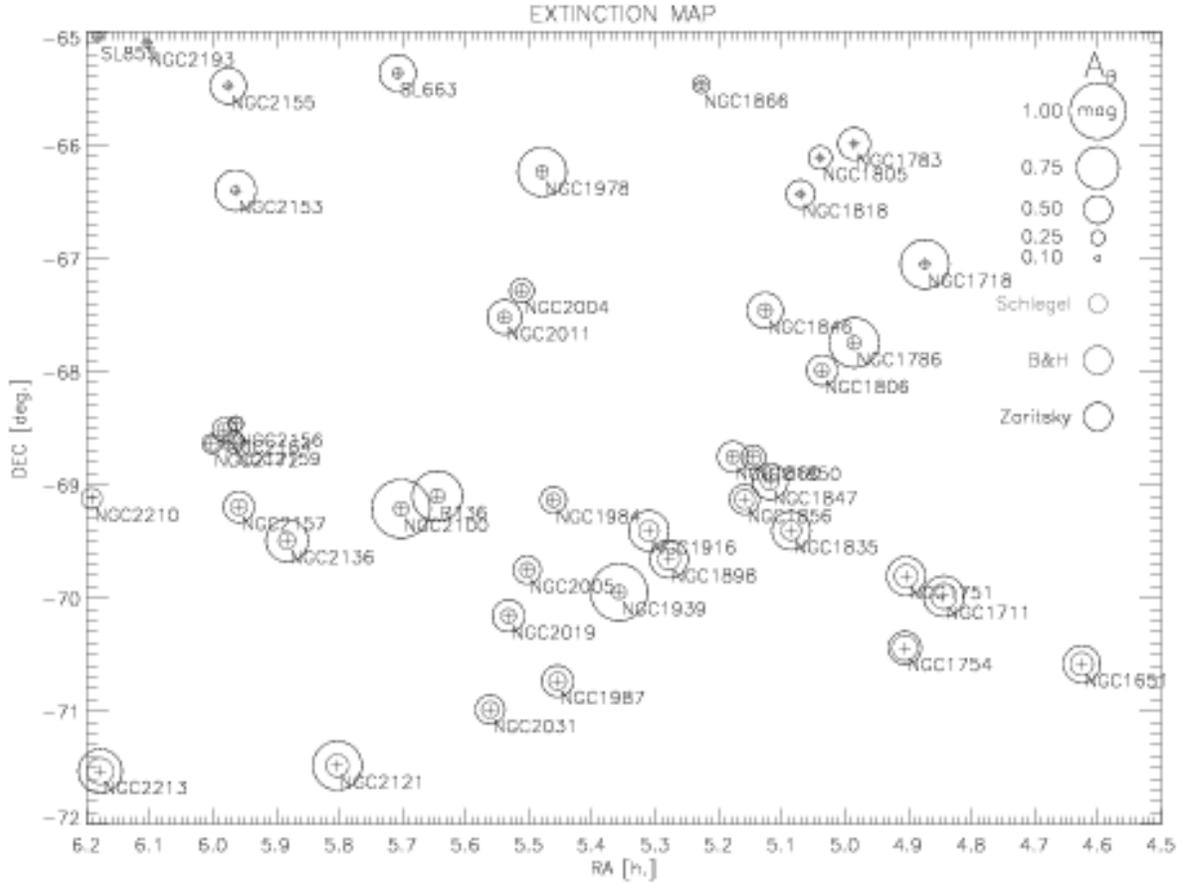}
\caption{Extinction map of the region of LMC where extinction estimations from
  the different studies are available for all objects in our sample. The values from \citeauthor{burstein82} and \citeauthor{zaritsky04} are plotted on the map and the extinction for LMC given by \citeauthor{schlegel98} ($A_B = 0.32 \pm 0.05$) is presented with the corresponding symbol size in the legend. }
\label{fig:ext}
\end{figure}

Extinction estimates based on HI emission and IRAS/DIRBE dust maps were
retrieved for all objects from NASA/IPAC Extragalactic Database (NED). The
values from MCPS were retrieved using the online tools available on the
webpage of the
project\footnote{http://ngala.as.arizona.edu/dennis/mcsurvey.html}. The
MCPS maps provide a statistical approach to the extinction: stellar
atmosphere models are fit to their measured {\it UBVI\/} magnitudes of all the
stars in an user defined search radius with good quality photometry and
good model fits \citep{zaritsky99}. The distribution of the extinction
values is built and the result of the mean extinction and
$\sigma_{ext}$ of the distribution is given. There are different options
available, but for the current work we chose to use the estimates based
on "cool" stars (T $\le$ $10^4$ K) in the search radius. Since cool stars
are more homogeneously distributed than hot stars, extinction
measurements from cool stars should provide a more 
representative estimate of the true extinction value. The extinction map
of the central region of the LMC, showing all the objects with estimates
available from the three studies is presented in Figure~\ref{fig:ext}.
Generally the values from \cite{burstein82} and \cite{schlegel98} are in
good agreement, but, in most cases, lower than those derived from
\cite{zaritsky04}. Using a search radius of 2\arcmin, which is slightly
larger than our largest photometry aperture, we achieve a robust estimate
of the extinction towards a specific object. 

\begin{figure}
\centering
\includegraphics[bb=-17 149 628 641,width=14cm]{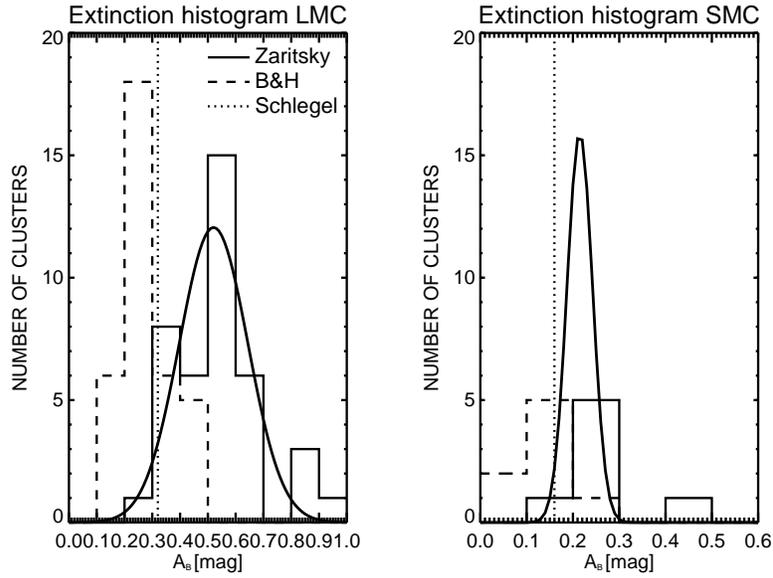}
\caption{Histograms of the extinction values from
  the studies, \citeauthor{zaritsky04} and \citeauthor{burstein82} for 
  the objects in the central regions of LMC and SMC. The data is presented in a
  similar way on the both panels. The Gaussian fit of the MCPS data is overplotted
  on each histogram. The values from \citeauthor{schlegel98} for both galaxies are denoted with vertical dotted lines.}
\label{fig:exthist}
\end{figure}

We use the extinction values based on the MCPS maps. 
In those cases where no MCPS data is available, we
adopt a typical extinction value, derived from all objects with extinction
estimates from MCPS. These mean values were derived by fitting the
extinction distribution for 40 LMC and 9 SMC clusters by a gaussian (see
Figure~\ref{fig:exthist}). The adopted values are: $A_B = 0.52 \pm 0.02$
and $A_V = 0.39 \pm 0.02$ for LMC, $A_B = 0.22 \pm 0.01$ and $A_V = 0.18 \pm
0.01$ for SMC. They are higher than the mean values from
\citeauthor{schlegel98}, which are $A_B = 0.32 \pm 0.05$ and $A_B = 0.16 \pm 0.03$
for LMC and SMC, respectively. 
This result is consistent with the fact that the extinction values for LMC and SMC listed in \cite{schlegel98} only provide lower limits: They only account for Galactic dust, whereas the dust in the Clouds is not taken into consideration.

We adopt the extinction law of \cite{bb1988}.

To account for the younger stellar population in objects in our sample with $log(age)<8.3$, we are using the extinction values based on all stars. This approach is providing a better estimate than "hot" star  (T $\le$ $10^4$ K) alone,  because it is reducing the influence of the relatively shallow MCPS $U$ band photometry.

%%%%%%%%%%%%%%%%%%%%%%%%%%%%%%%%%%%%%
\subsection{Notes on individual objects}
\label{notes}
There are a few cases in which the data reduction and photometry
differed slightly from the procedure described above. Additional
remarks for these clusters are provided below. 

{\bf NGC~121} There is a relatively bright star located $\sim\!60$\arcsec\ 
W from the center of the cluster. An inspection of 2MASS and optical images 
(from SIMBAD\footnote{http://simbad.u-strasbg.fr/}) showed that it is most 
likely a foreground star superposed 
on the cluster. Its magnitudes from ALLSTAR output files are $J = 10.78$, 
$H = 10.22$ and $K_s=10.11$. The extinction correction values are taken 
from \cite{burstein82} and \cite{schlegel98} (we discuss extinction
corrections in detail in 
Section~\ref{ext}). They are $A_B=0.15$ and 0.16, respectively, and we
use their mean in this analysis. The 
measured colors are: $(J-H)_0=0.55$, $(H-K_s)_0=0.104$, $(J-K_s)_0=0.65$. 
Comparing with the results of \cite{ducati01} for intrinsic NIR colors of stars 
shows that its colors are consistent with a K main-sequence, G giant or 
supergiant star. In the latter case, rough derivation of the absolute 
visual magnitude at the distance of SMC gives $M_V\approx-7.5$ mag. This
is consistent with the star being a supergiant, but it is not likely that 
such an object is found in a $\sim\!12$ Gyr old star cluster. Therefore, 
we assume that the star belongs to the foreground and subtract it prior to 
the final cluster photometry.

{\bf NGC~339} This object is relatively close to the edge of the Atlas Image.
The radius of the largest aperture used for photometric measurements is 90\arcsec. 
In all cases, when the maximal size of the apertures was smaller 
than the typical value of 200\arcsec,  the last entries for the particular object 
in Table~\ref{tab:photsmc} or \ref{tab:photlmc} are the magnitudes measured in the 
largest apertures used.

{\bf NGC~419} There is a bright star in the aperture area 
$\sim\!90$\arcsec\ SSE from the center used for photometry. It is clearly 
visible on the optical frames retrieved from SIMBAD. Given its colors of 
$J-K_s = 0.410$ and $H-K_s = 0.004$ we conclude that it is most likely a 
foreground object and subtract it prior to the final measurements.

{\bf NGC~458} This young cluster is barely visible in the 2MASS frames, in 
particular in $H$ and $K_s$ where the infrared sky background is 
significantly higher. The curves of growth in these bands start to 
decline for aperture diameters larger than 80\arcsec. We provide 
integrated magnitudes only up to 90\arcsec\ aperture radius, due to the 
proximity of this cluster to the edge of the Atlas Image, but the results 
for the largest radius must be treated with caution.

{\bf Lindsay~1} The cluster was split almost equally between two sets of 
frames. To derive its integrated magnitudes we could not use the usual
centering routine and the center was derived "by eye", accounting for the 
appearance and position of the cluster on each image. The fluxes from the 
two halves were measured independently, summed together, and converted
into magnitude values. The frames originate from the same scan, acquired 
on Aug. 8 1998 between 07:03:03.00 and 07:08:51.00 UT. The regions of the 
sky on these images were observed at 07:03:43.51 and 07:04:01:38 UT one 
after another. Hence, we averaged the zero points of the frames and used 
that value for the derivation of the magnitudes. The mean zero points were 
20.8522, 20.4090 and 19.8725 in $J$,$H$ and $K_s$, respectively. The 
errors of the photometry were estimated in accordance to the described
procedure.

{\bf NGC~1711} There is a chain of so-called persistence artifacts in close 
vicinity to the cluster. These features most likely originate from a 
bright star outside of the current Atlas Image. Two of them affect 
the photometry, are located at distances $\sim$47\arcsec\ and $\sim$66\arcsec\ 
from the center. They are well outside of the unresolved cluster component, and were 
measured independently on the residual frames. Their flux was subtracted 
from the affected apertures before calculation of the magnitude values. 
The rest of the artifacts were avoided by specifying a larger 
radius for the background annulus. 
The resulting errors for this cluster were estimated by taking into account 
the effect of the artifact removal.

{\bf NGC~1754} A bright star is present $\sim\!35$\arcsec\ SE from the 
center of the cluster, also clearly visible on the visual frames from 
SIMBAD. The magnitude values derived from our PSF photometry are $J$=9.31, 
$H$=8,79, $K_s$=8.68 mag and the resulting colors: $J-H$ = 0.52, $H-K_s$= 
0.11, $J-K_s$ = 0.63. This case is similar to NGC~121 in the SMC. The 
extinction value towards that object was retrieved from the reddening estimator for
LMC on the webpage of MCPS $A_V$ = 0.4. The corrected colors of the  star are then
$(J-H)_0=0.48$, $(H-K_s)_0=0.08$ and $(J-K_s)_0=0.56$. These colors
best match those of a G4 giant from \cite{ducati01}. The measured  absolute visual 
magnitude is not compatible with the predictions for a G4 giant. The estimated age 
of the cluster ($\sim$ 15Gyr) is rules out the possibility that the star is a 
supergiant. Most likely it is a foreground star
and we excluded it from the measurements of the total cluster luminosity. 

{\bf NGC~2136} This is one of the most interesting objects in our sample. The cluster is
$\sim$ 100 Myr old and there is a "satellite cluster" clearly visible on
both visual and near-infrared images $\sim$ 80\arcsec \  from the central
position derived in the  present work. The difference between the coordinates
retrieved from SIMBAD and the actual center is 166 pixels on the 2MASS  $J$
frame. The object is also off-centered on the optical frame downloaded from the
same database as the coordinates. We conclude that the most
probable cause of this discrepancy is a mistake in the coordinates listed in
SIMBAD. They are given in Table~\ref{tab:smc}, and the values derived for the
centering in the present work are listed in  Table~\ref{tab:photlmc}. 

There are two bright stars in the set of apertures used to built the curve-of-growth. 
An analysis similar to the case of the stars in the field of NGC~121
and NGC~1754 led us to the conclusion
that the  absolute visual magnitudes 
differ from these expected for luminous stars of these spectral
types. We chose at  that point to exclude them from
the final photometry. 

{\bf NGC~2153} This object was situated too close to the frame edges on all the
sets of Atlas Images available. We chose those with the best possible
location, but the largest aperture radius is still only 40\arcsec\
before running into the frame edge. However, the
cluster is compact and even in that significantly smaller aperture set
(compared to that typically used in this work) the curves-of-growth indicate
sampling of the entire flux from the object. In fact there is some decline
observed in $H$ for aperture radii larger than 20\arcsec. The most likely
explanation is local variation in the background, typical for the $H$ band. Due to the
position of the object the background levels were estimated in a region of the
sky close to the cluster. We
present the results for the complete set of apertures, but the values of $H$
magnitudes must be treated with caution for radii exceeding 20\arcsec. 

{\bf SL~842} This compact cluster is barely detected by 2MASS. Photometry is
performed with the entire set of apertures, but the results become highly
unreliable for aperture radii exceeding 30\arcsec. This aperture
appears to encompass the 
measurable flux from the object and these are the results listed in
Table~\ref{tab:photlmc}. 

{\bf SL855} The cluster is barely visible on the Atlas Images. The photometry
was initially  performed with the entire set of apertures. The shape
of the curves-of-growth  
and inspection of the frames led us adopt a more
conservative approach, and we list only the results to aperture radius 10\arcsec.

{\bf ESO121-003}  The cluster is faint and extended. It is detected by
2MASS, but the curves of growth are noisy.

%%%%%%%%%%%%%%%%%%%%%%%%%%%%%%%%%%%%%
\section{Results}
\begin{figure}
\centering
\includegraphics[bb=-28 56 609 693,width=12cm]{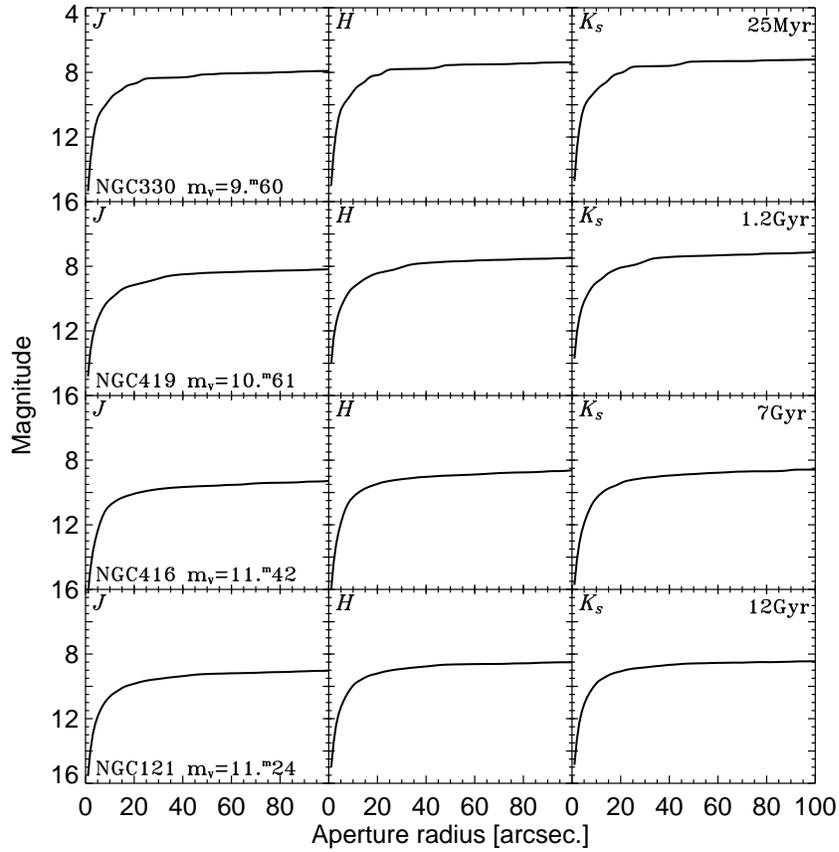}
\caption{Curves of growth for four SMC clusters from our sample. The visual
  magnitudes are taken from \cite{alcaino78}. The electrophotometry for all
  clusters in his study is done by using a 62\arcsec aperture. The 2MASS
  $J$,$H$ and $K_s$ Atlas Images of these clusters with size representable for
  the diameter of our largest aperture are shown on Figure 1.} 
\label{fig:cg4smc}
\end{figure}

\begin{figure}
\centering
\includegraphics[bb=-28 56 609 693,width=12cm]{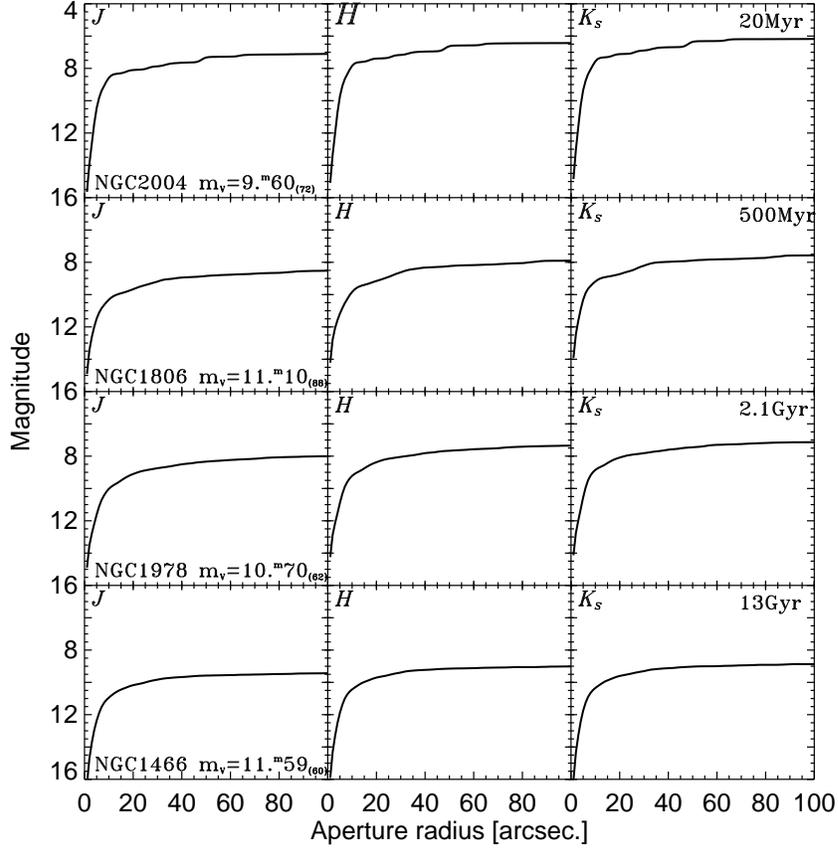}
\caption{Curves of growth for four LMC clusters from our sample. The visual
  magnitudes are taken from the compilation of \cite{bica96}. Those data is
  originating from numerous sources and is not homogeneous by means of
  detectors and aperture sizes. The apertures used for the measurements of the
  magnitudes cited on the plots are given in parenthesis.The  2MASS $J$,$H$ and
  $K_s$ Atlas Images of these clusters with size representable for the diameter
  of our largest aperture are shown on Figure 2.}
\label{fig:cg4lmc}
\end{figure}

The results from the integrated 2MASS photometry of the entire Magellanic 
Clouds cluster sample are presented in Tables~\ref{tab:photsmc} and
\ref{tab:photlmc}. (The entire 
tables are available in the electronic version of the journal.) 
A set of typical NIR $J$, $H$, and $K_s$ curves of growth of four SMC
clusters, ranging in age from $\sim$\,25 Myr to $\sim$\,12 Gyr is given on
Figure~\ref{fig:cg4smc}. A closer look at the curve of the youngest
cluster NGC~330 reveals well visible ''bumps''. These are bright stars
contributing to the total light; these are common for young and some of 
the intermediate-age clusters, and are most likely massive young 
supergiants and carbon stars, which  emit significant amounts of 
light in the NIR. Figure~\ref{fig:cg4lmc} presents curves of growth for 
LMC clusters also covering a representative age range. The 
corresponding images of these objects are presented in 
Figures~\ref{fig:smc} and \ref{fig:lmc} for the SMC and LMC, respectively.

\begin{figure}
\centering
\includegraphics[bb=28 28 594 708,width=10cm]{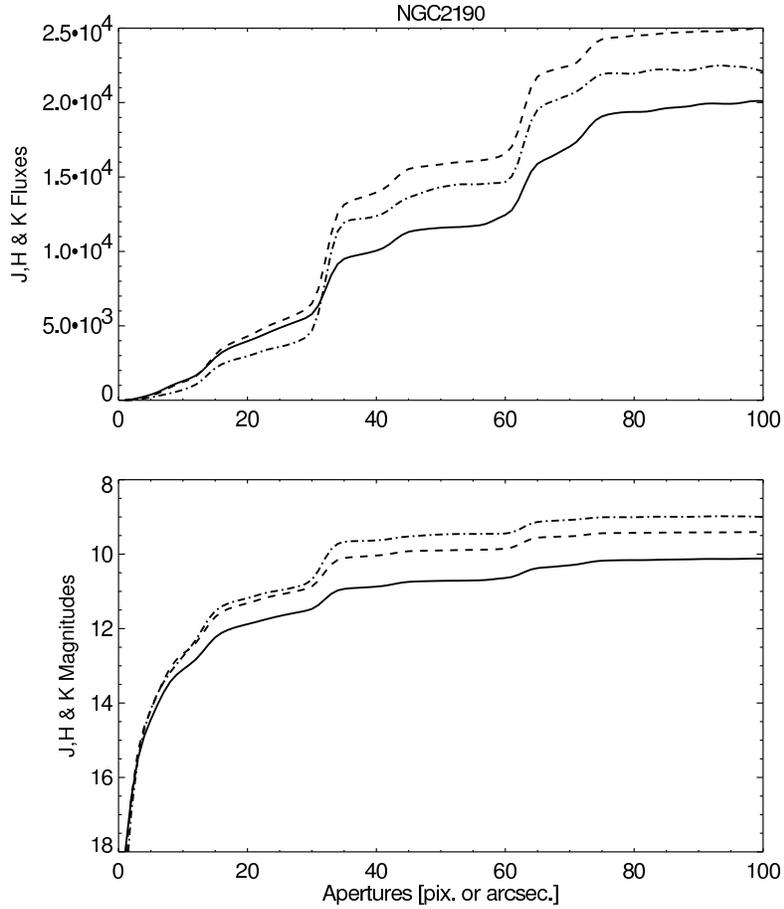}
\caption{Curves of growth in flux units (upper pannel) and magnitudes (lower
  pannel) for the LMC cluster NGC2190 affected by carbon stars. Note the
  features around r $\sim$ 30\arcsec and 60\arcsec and the steep increase of the flux when the carbon stars are entering into the aperture.} 
\label{fig:n2190}
\end{figure}

The carbon stars present in some intermediate-age clusters 
%PP - added the following reference to Frogel et al.1990 after the final referee comments
are easily distinguished by their colors and luminosity as mentioned by \cite{frogel90}.
%ended here
They also affect the curve of growth in a 
typical way, leaving a "fingerprint" of their presence. A good example is 
the LMC cluster NGC~2190. The curves of growth for this cluster cover the 
carbon star KDMK6996 \citep{kontizas01} and another candidate carbon star 
closer to the cluster center (see Figure~\ref{fig:n2190}). The features at 
$\sim\!30$\arcsec\ and 60\arcsec\ radii are caused by carbon stars 
entering the aperture. Note the steeper increase of the curve of growth for
$K_s$ compared to those for $J$ and $H$, 
and the corresponding features in the flux curves. The carbon star 
identification is based on the magnitudes and colors from their PSF 
photometry. It is easy to detect the red colors of these objects in the 
NIR passbands. Their intrinsic colors are expected to be 
$(J-H)_0\approx1$, $(H-K)_0\approx1$ and $(J-K)_0\approx2$ 
\citep{ducati01}. 

%%%%%%%%%%%%%%%%%%%%%%%%%%%%%%%%%%%%%
\subsection{Comparison with previous studies}
We compare our results with the work of \cite{persson83} in this section.
There are 52 objects in common between our study and their paper: 10 SMC
and 42 LMC clusters, respectively. The data in that early work was
gathered using three different photo tubes and an InSb detector system,
mounted on three different telescopes: The 1-m Swope and 2.5-m du Pont
telescopes of the Las Campanas Observatory and the 0.9-m CTIO telescope.
The observations of Magellanic Cloud clusters were presented in $J$,$H$
and $K$ filters of the California Institute of Technology infrared
photometric system (CIT) \citep[for details see][]{frogel78}.

\begin{figure}
\centering
\includegraphics[bb=88 231 473 610,width=10cm]{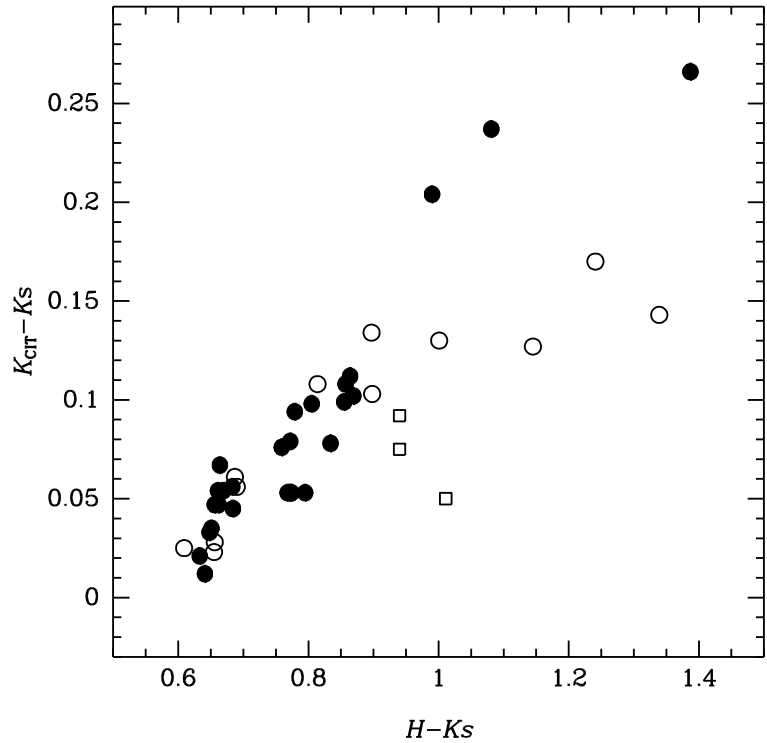}
\caption{Relation between $H-Ks$ and $K_{\rm CIT}-K_s$ for flux-calibrated 
spectra of late-type stars taken from \cite{lancon92}. Filled
  circles represent data of giants of spectral types G5III to M8III,
  open circles represent data of supergiants of spectral types G2I --
  M7I, and open squares represent data of carbon stars. Note
  that any difference in calibration from $K_{\rm CIT}$ to $K_s$ between
  the different spectral types is insignificant for $H-K_s < 0.9$.}
\label{fig:HmKs_KjmKs}
\end{figure}

There are several issues that complicate a direct comparison of the
obtained results in the two works: {\it (i)\/} Due to the use of an
iris diaphragm at the du Pont and CTIO telescopes at the time, the
aperture diameters were only known to $\pm1$\arcsec\
\citep{persson83}. This could lead to uncertainties in the cluster
magnitudes and colors. {\it (ii)\/} Another serious problem we became
aware of during a series of experiments is related to the centering of
the cluster. In many cases the diaphragm apertures used by Persson et
al.\ appear to cover the brightest part of the cluster, because their
strategy was to maximize the flux through the aperture. This however
leaves this technique vulnerable to the effect of stochastic
fluctuations of the observed stellar population, in particular for
young clusters or clusters that are contaminated by bright
stars. Extended clusters without a clear central peak are also
difficult to center using this technique. {\it (iii)\/} The
cross-calibration between the CIT and 2MASS photometric system was
only based on three stars with $(H\!-\!K)_{\rm CIT} > 0.5$
\citep{carpenter01}. However, one might expect differences in
calibration from CIT $K$ to 2MASS $K_s$ for late-type giants
vs.\ supergiants or carbon stars (i.e., stars with $(H\!-\!K)_{\rm
  CIT} > 0.5$), since the latter two have stronger CO bandhead absorption
features (which affects $K$ much more than it does $K_s$). As this may
be relevant for intermediate-age star clusters whose near-IR colors
are dominated by light from AGB stars, we tested the significance of
this effect by using the {\sc synphot} package within IRAF/STSDAS
along with $H$-- and $K$-band spectra of late-type giants,
supergiants, and carbon stars taken by \cite{lancon92} and filter
throughput curves taken from \cite{cutri03}. As
Fig.~\ref{fig:HmKs_KjmKs} shows, the offset between $K$ and $Ks$ for
the different types of late-type stars only starts being significant
redward of $H\!-\!Ks \sim 0.9$. Since the clusters in our sample all
have $H\!-\!Ks < 0.9$ (cf.\ Tables~\ref{tab:photsmc} and
\ref{tab:photlmc}), we conclude that color term differences between supergiants and carbon stars do not significantly influence the integrated-light photometry of our clusters, and hence the cross-calibrations of
\cite{carpenter01} should be adequate for our purposes.

\begin{figure}
\centering
\includegraphics[bb=34 57 590 728,width=14cm]{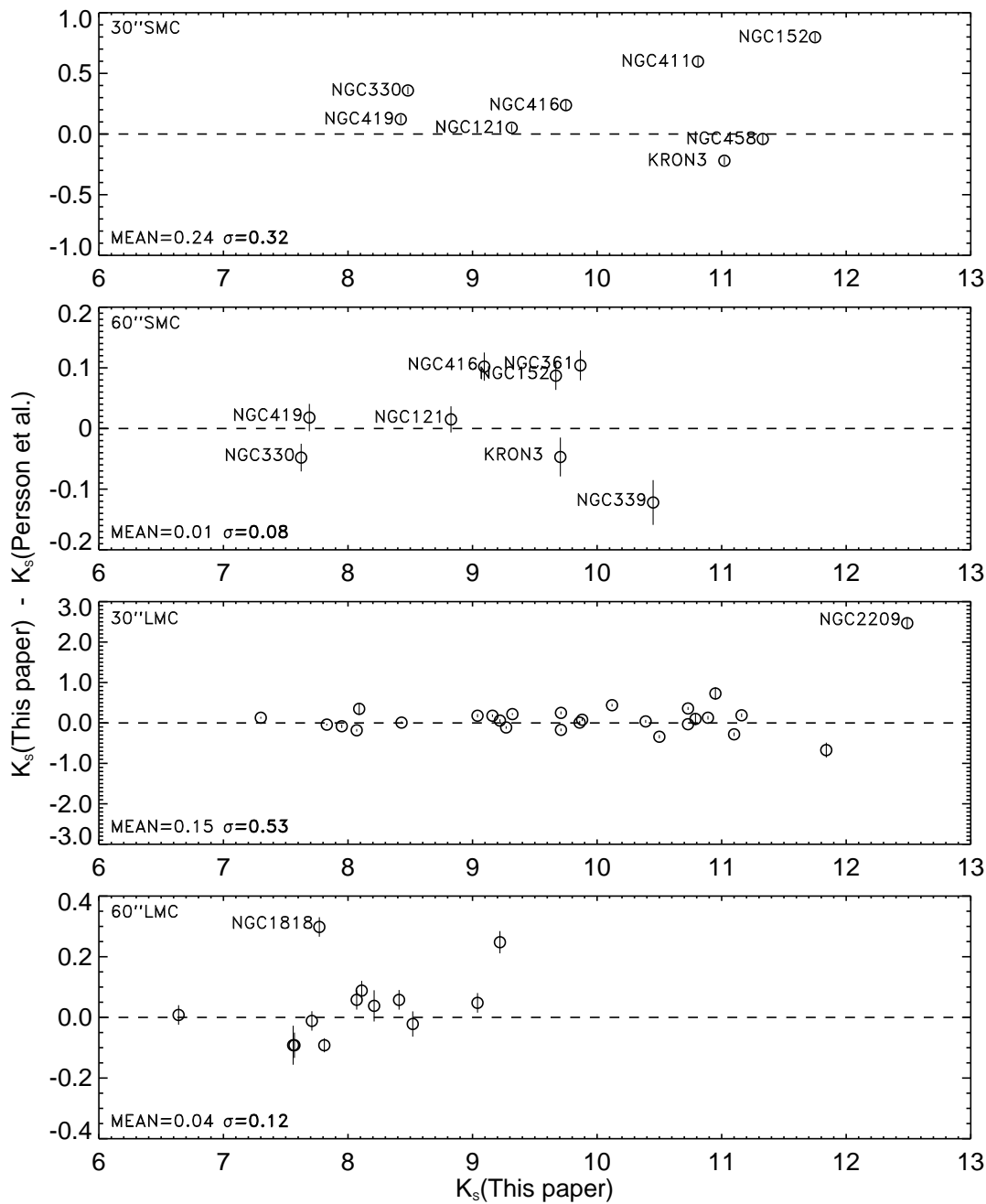}
\caption{Comparison of our 2MASS $K_s$ band photometry for LMC and SMC clusters with the results of \cite{persson83}. The dashed line is the one-to-one relation.} 
\label{fig:compk}
\end{figure}

\begin{figure}
\centering
\includegraphics[bb=45 130 557 620,width=12cm]{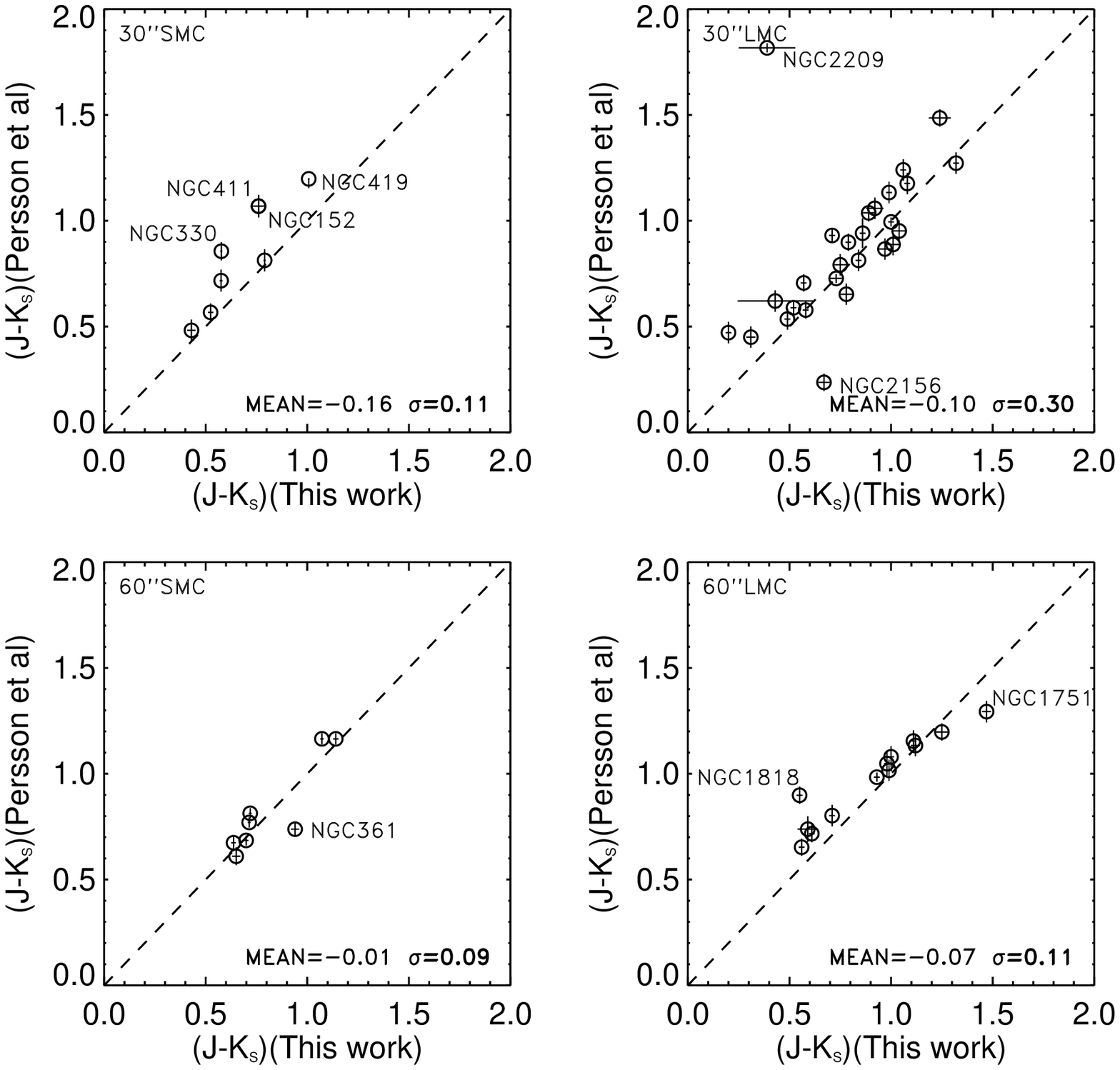}
\caption{Comparison of our 2MASS $J-K$ colors for LMC and SMC clusters with \cite{persson83}. The dashed line is the one-to-one relation.}
\label{fig:compjk}
\end{figure}

\begin{figure}
\centering
\includegraphics[bb=45 130 557 620,width=12cm]{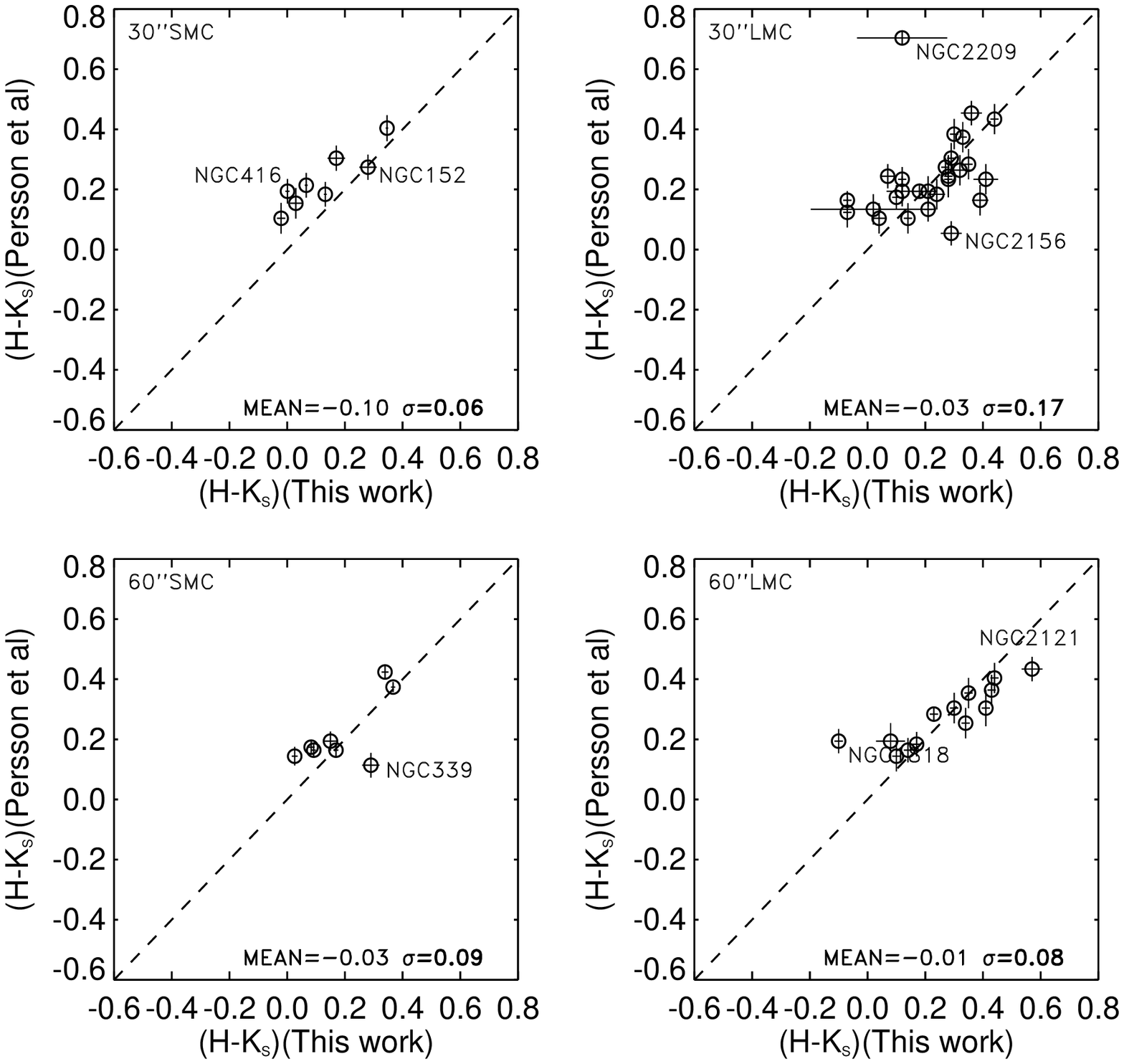}
\caption{Comparison of our 2MASS $H-K$ colors for LMC and SMC clusters with \cite{persson83}. The dashed line is the one-to-one relation.}
\label{fig:comphk}
\end{figure}

The $K$ CIT magnitude values as well as ($J-K$) and ($H-K$) colors
from \cite{persson83} were thus
converted into the 2MASS system by using the transformation equations
derived by \cite{carpenter01}\footnote{The most recent update of the
  transformations is available online at:
  http://www.astro.caltech.edu/~jmc/2mass/v3/transformations/}. The
comparison plots between $K_s$ magnitudes and the colors for the SMC
clusters in common between our work and \citeauthor{persson83} are
presented in Figures~ \ref{fig:compk}--\ref{fig:comphk}. The mean offsets between the $K_s$ magnitude values from the two studies are $0.13$ ($\sigma = 0.26$) and $0.11$ ($\sigma = 0.44$) for SMC and LMC, respectively. For $(J-K_s)$ color we calculated mean offset $ -0.08$($\sigma = 0.13$) and $-0.08$($\sigma = 0.25$) for the SMC and LMC. The $(H-K_s)$ mean offsets for the two galaxies are $-0.06$($\sigma = 0.08$) and $-0.09$($\sigma = 0.16$).

The largest differences appear in the case of NGC~152, situated in the SMC and
NGC~2209 in the LMC. If we can explain the nature of these
discrepancies it is plausible to assume that it is possible to explain
the smaller offsets arising for the rest of the objects in the
sample as well. 

\begin{figure}
\centering
\includegraphics[bb=14 14 363 535,width=12cm]{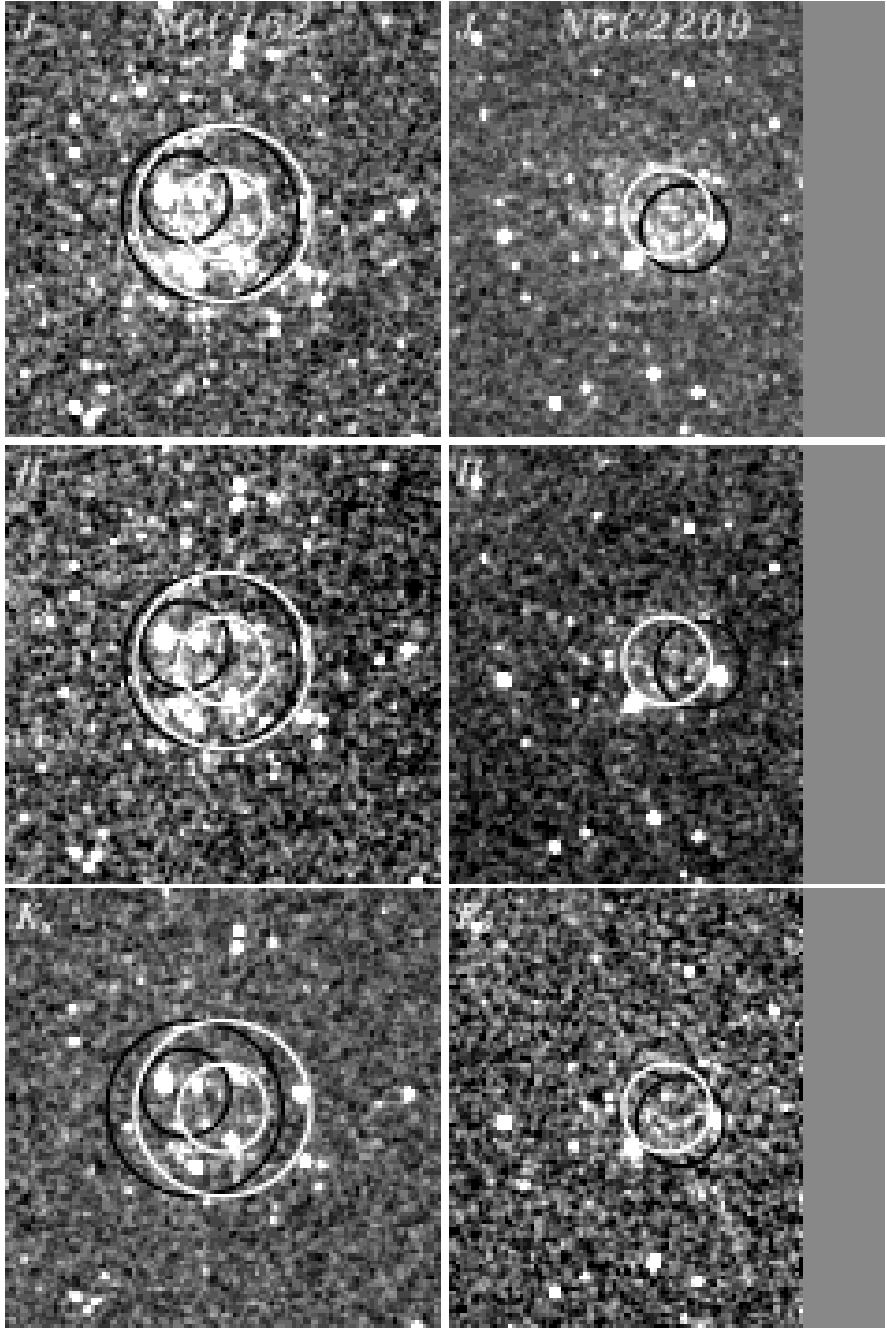}
\caption{2MASS images of NGC152 and NGC2209. North is up and East is to the
  left. The images are centered on the positions of our apertures and the size
  of each one is 200\arcsec square. The grey part on the left set of pannels
  indicate that the object is close to the edge of the Atlas Image. The
  reproduced apertures of  \cite{persson83} are plotted in black, and the
  corresponding apertures from our work are shown in white.} 
\label{fig:n152}
\end{figure}

To investigate this in more detail we plot the $J$, $H$ and $K_s$
frames of NGC~152 on the right side of Figure~\ref{fig:n152}. The
images are centered on the cluster position derived in this paper (as
explained above). To simulate the measurements of \cite{persson83}
who maximized the count rate received through their single-channel
detector, we used an aperture of their size and let its center drift
across a 40\arcsec\ $\times$ 40\arcsec\ subimage located around the
center position of our apertures. The step size was 1 pixel (or
arcsec) and after the row or column was completed, the aperture center
moved to the next one until the entire section was scanned. The
measurement with the maximal flux value is assumed to be the center of
that aperture in \cite{persson83}. These simulations were done for
each passband independently. The results are listed in
Table~\ref{tab:comp152} and Table~\ref{tab:comp2209}. For illustration
purposes, we plot the apertures used in our study and the reproduction
of those used by \citeauthor{persson83} together in
Figure~\ref{fig:n152}. 
NGC~152 is an intermediate-age SMC cluster, and 
there are several bright red stars which dominate the flux in the 
near-infrared. (Their presence was also noted by
\citeauthor{persson83}) 
The %shallow 2MASS imaging
faintness of the cluster, the bright stars and the extended nature of
the cluster render the centering extremely hard for single-channel photometry. 
The problem is most serious for the smaller aperture, but there is a
better agreement for the bigger one. The results of the comparison are
shown in Table~\ref{tab:comp152}. 

\begin{figure}
\centering
\includegraphics[bb=56 56 509 736,width=14cm]{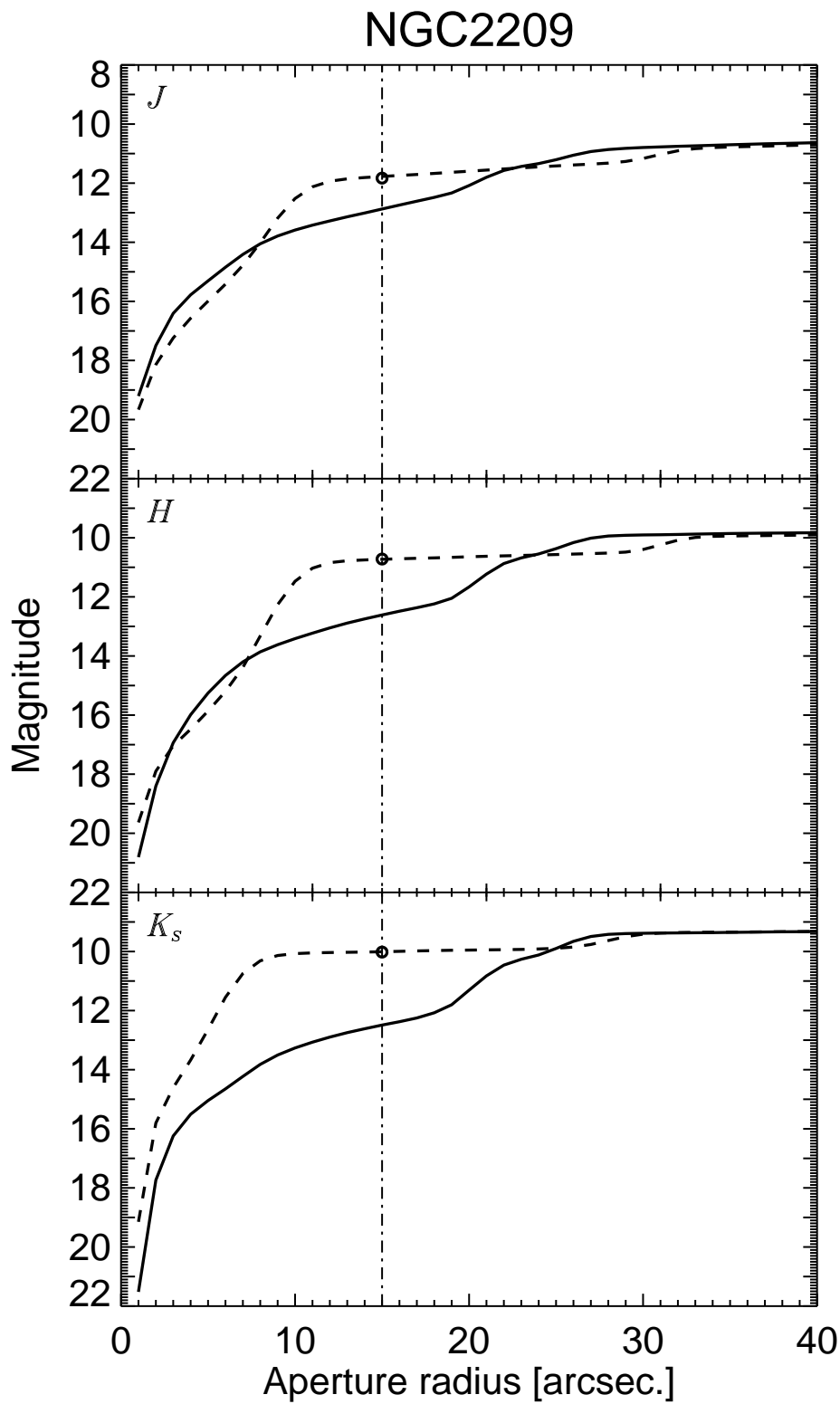}
\caption{$J$, $H$ and $K_s$ curves of growth for NGC2209. The solid lines are illustrating  our photometry and the dashed lines are standing for the photometry performed with the reproduced centering of \cite{persson83}. Dot and dashes are showing the 30\arcsec aperture diameter. The points are presenting magnitude values from \citeauthor{persson83} and their error bars are compatible with the point size. Note the good agreement between the measurements for larger apertures, showing that the large discrepancies in D$=30\arcsec$ are due mostly to the different centering.}
\label{fig:ngc2209cg}
\end{figure}

NGC~2209 is the most extreme example of differences between our study
and that of \cite{persson83}, amounting to 2.5 magnitudes in the
$K$-band. The 2MASS images of this cluster are shown in the left three
panels of Figure~\ref{fig:n152}. The object is a faint cluster with an
age of $\sim$\,1 Gyr. There are two bright stars dominating the flux,
which are carbon stars identified as W46 and W50 by
\cite{walker71}. \citeauthor{persson83} note that they may affect the
centering of their aperture and therefore likely the results of their
photometry, which was performed with a single 30\arcsec\ aperture. The
offsets of the Persson's aperture centers, reproduced by maximal flux
experiments is between 10.5 and 17.5 pixels away from our position,
depending on the passband used. 
%PG% let's join the previous and the following paragraphs. 
% Both are on NGC 2209. 
The magnitude and color values listed in Table 2 of \cite{persson83}
and the curves of growth from our work (presented in
Figure~\ref{fig:ngc2209cg}) suggest that the flux of the carbon stars
in NGC~2209 
affects the total cluster magnitudes of \citeauthor{persson83} 
%PG% Entered the following (important I think for the reader to understand)
To test our simulated position of Persson et al.'s aperture, we
converted the magnitudes of \cite{persson83} to the 2MASS system and compared 
them with the corresponding values from our work after recentering our
aperture on the simulated position used by Persson et al. The results   
are presented in
Table~\ref{tab:comp2209}. There is good agreement between $K_s$
magnitudes and $H-K_s$ colors, $J-K_s$ is a little off, but there
still is good agreement at the 3$\sigma$ level. The most probable
reason is a slight difference in the $J$ magnitude values. This is not
surprising taking into account that the $J$ band magnitudes could be 
affected by 
rapid variations of the water vapor content in the atmosphere. In this
particular case the differences between our results and the photometry
of \citeauthor{persson83}\ are mainly caused by centering problems. We are
taking into account only the photometric uncertainty and the errors of
the 2MASS zero points in the analysis above. Due to the presence of
several relatively bright stars in the background field and the low
signal from the cluster, the errors associated with the stochastic
fluctuations in the stellar background are quite high. If we take them
into account, there is much better agreement between the magnitudes
and colors reproduced by our experiment and the values of
\cite{persson83}. 

In general, we were able to achieve agreement between our results and
those of \citeauthor{persson83} by assuming a center location for
their measurements which is significantly off the `true' center of the
cluster in question. The centering discrepancies are smaller for the
larger apertures, but still large enough to alter the total magnitudes
significantly. 

Figure~\ref{fig:compk} is perhaps the best illustration of the effects
mentioned above. The total magnitudes of the clusters in
\cite{persson83} for the 30\arcsec\ aperture compared to the data in
the present paper are underestimated for 6 of 8 objects and the most
probable explanation is centering of the aperture over the brightest
part of the cluster population. There is an overestimate in the case
of KRON~3, which is a little surprising taking into account the
compact nature and the shape of the object. On the other hand a more
careful inspection shows that if the aperture is placed on the
geometrical center of KRON~3 it is not sampling the most luminous part
of the stellar population. In general we have good agreement for the
compact and bright clusters, residing in regions of relatively
homogeneous foreground or Magellanic clouds stellar population
(e.g. NGC419, NGC121), especially for the larger 60\arcsec\
aperture. This is also the case for NGC~458, a faint cluster, measured
by \citeauthor{persson83} only in a single 30\arcsec\ aperture. The
values of $J-K_s$ and $H-K_s$ colors of the clusters for the smaller
30\arcsec\ aperture are systematically higher in
\citeauthor{persson83}. The observed trend is also consistent with the
expected results from the flux-maximization. 

\cite{kyeong2003} presented 2-dimensional NIR imaging of a smaller sample of 28 LMC 
clusters. Their observations were conducted in December 1996 with CASPIR (Cryogenic Array 
Spectrometer/Imager) instrument at the 2.3m telescope of the Siding Spring Observatory. 
The clusters were observed in the $JHK$ passbands of the SAAO NIR system. The 
flux from the background fields was subtracted from the flux measured in the photometry 
aperture. The authors took advantage from their imaging data to determine the centers of the 
objects by visual inspection and used them to measure the integrated magnitudes of the clusters 
in 11 concentric apertures. Unfortunately the center positions were never published, so it is 
impossible to provide a detailed comparison between our photometry and the values in the earlier 
work. We transformed the values of the $J$,$H$ and $K$ total magnitudes for their largest apperture  (D=100\arcsec) into 2MASS magnitudes for the 22 objects in common, using the work of \cite{carpenter01}. The comparison between the datasets showed mean offsets (by means of difference between our magnitudes and the values of \citeauthor{kyeong2003}) : $-0.10\pm0.05$, $-0.06\pm0.04$ and $0.00\pm0.03$ in $J, H$ and $K_S$ respectively.  An inspection 
of the observing log published in their Table 2 revealed notes about non-photometric conditions 
concerning clusters observed during two of the nights.  This is a possible explanation for the larger differences 
between our photometry and the results of \citeauthor{kyeong2003} in $J$ and $H$ bands which
are much more affected by rapid changes of the atmospheric transition and water vapor content.

%%%%%%%%%%%%%%%%%%%%%%%%%%%%%%%%%%%%%
\subsection{Comparison with 2MASS Extended Source Catalogue} 
\begin{figure}
\centering
\includegraphics[bb= -28 85 609 722,width=16cm]{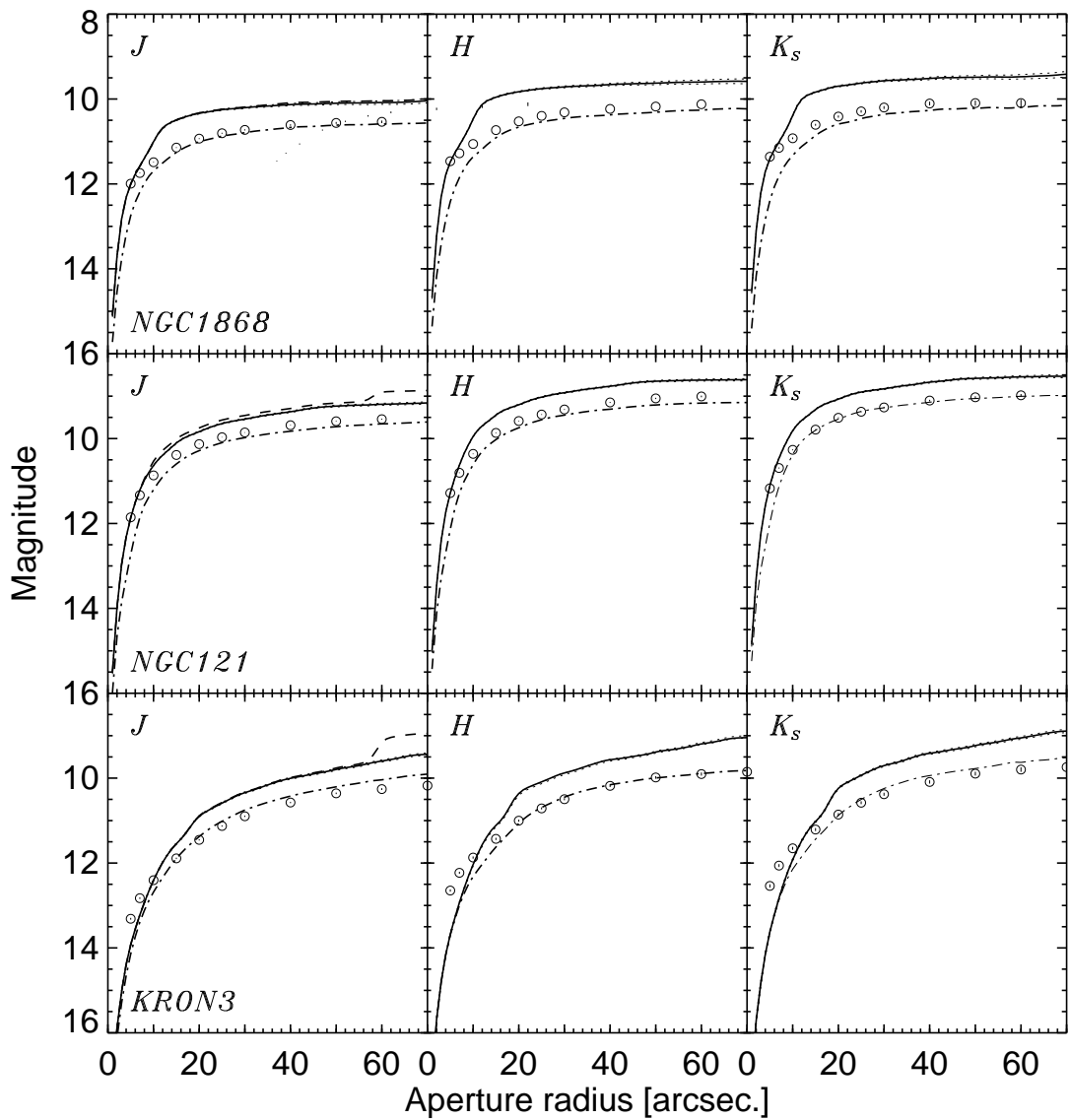}
\caption{Comparison between our photometry (solid lines) and 2MASS XSC values (circles). The errors are compatible to the thickness of the lines. The total signal from the clusters is shown for clarity only in $J$. XSC values are in much better agreement with the unresolved component (dots and dashes). The differences in the case of KRON3 could be attributed to the centering of XSC apertures on the peak $J$ pixel.}
\label{fig:compxsc}
\end{figure}

The Extended Source Catalogue (hereafter XSC) processor in the 2MASS pipeline was designed to provide a flux measurement of the diffuse light of extended sources such as distant galaxies. As described in \cite{jarrett2000}, the XSC processor masks out point sources and substitutes the flux in the masked pixels with the surface brightness of the underlying diffuse light. When applied to star clusters in the Magellanic Clouds, one can therefore expect the XSC processor to eliminate stars that are actually genuine members of the star clusters, some of which contribute significantly to the total flux. As shown in Fig.~\ref{fig:compxsc}, we indeed find that the XSC magnitudes (which are given for 11 concentric circular apertures) are in very good agreement with our photometry of the {\it unresolved\/} component of the clusters. However, as the {\it total\/} magnitudes of the clusters are significantly brighter than this, we discourage use of the XSC catalog for partially resolved targets such as those considered here.

%%%%%%%%%%%%%%%%%%%%%%%%%%%%%%%%%%%%%
\section{Summary and Conclusions}
We present a highly uniform dataset of integrated $J$, $H$ and $K_s$
magnitudes for 75 star clusters in the Magellanic Clouds, using 
2MASS survey data. There are reliable age and metallicity estimates
available in the literature for the vast majority of the objects in
the sample. This is the first extensive NIR survey 
of the clusters in these galaxies since the single-channel photometry
of \cite{persson83}. Comparing the results of their photometry with the
magnitudes from our infrared array curve-of-growth measurements, we find
significant differences for some objects, which we can reproduce as being due to centering
problems in the early \citeauthor{persson83} study.

Keeping in mind that the \cite{persson83} results were used to calibrate some 
of the most recent SSP models \citep{bc03, maraston05}, we suggest that the
photometry derived in the present work be used to calibrate and
improve the existing and future SSP models in the near-IR part of the spectrum.
We intend to perform a detailed comparison with the predictions of a set of simple stellar population 
models in a forthcoming paper, utilizing new $VRI$ optical data from \citet[][]{goudfrooij06}.

%%%%%%%%%%%%%%%%%%%%%%%%%%%%%%%%%%%%%
\paragraph*{Acknowledgments} 
The authors would like to thank to the anonymous referee for the helpful 
suggestions and comments that improved the paper. 
The authors are thankful to Dennis Zaritsky and his collaborators for the
kind and quick response of the inquiry about the MCPS extinction estimation
tool and providing access to their online 
utility. %\footnote{http://ngala.as.arizona.edu/dennis/smcext.html}. 
This publication makes use of data products from the Two Micron
All-Sky Survey, which is a joint project of the University of 
Massachusetts and the Infrared Processing and Analysis
Center/California Institute of 
Technology, funded by the National Aeronautics and Space Administration
and the National Science Foundation. This research has made use of the
NASA/IPAC Extragalactic Database (NED) which is operated by the Jet
Propulsion Laboratory, California Institute of Technology, under contract
with the National Aeronautics and Space Administration. 
Support for this work was provided in part by
NASA through a Spitzer Space Telescope Program, through a contract
issued by the Jet Propulsion Laboratory, California Institute of
Technology under a contract with NASA. Support for this work was also 
provided in part by the STScI Director's Discretionary
Research Fund.

%%%%%%%%%%%%%%%%%%%%%%%%%%%%%%%%%%%%%

\clearpage

\begin{deluxetable}{lccrrrlcrlrrr}
\tabletypesize{\small}
\tablenum{1}
\tablecolumns{12}
\tablecaption{SMC cluster sample.}
\tablehead{
\multicolumn{1}{l}{ID}&\colhead{$\alpha_{2000}$\tablenotemark{a}}&\colhead{$\delta_{2000}$\tablenotemark{a}}&
\multicolumn{4}{c}{Age, Errors \& Ref\tablenotemark{b}}&\multicolumn{3}{c}{[Fe/H], Err \& Ref}&
\colhead{$A_v$\tablenotemark{c}}&\colhead{$A_v$ err\tablenotemark{c}}}
\startdata
NGC121      &   00:26:49  &--71:32:10  & 10.08 & +0.05   &--0.05   &  7  & --1.71   &$\pm$ 0.10 &         7       & 0.18 & 0.02\\ 
NGC152      &   00:32:56  &--73:06:59  &  9.15 & +0.06   &--0.07   &  3  & --0.94   &$\pm$ 0.15 &         3       & 0.19 & 0.02\\ 
NGC176      &   00:35:59  &--73:09:57  &  8.30 & +0.30   &--0.30   &  6;8& --0.6\phn&\nodata    &         8      & 0.24 & 0.03\\
NGC330      &   00:56:20  &--72:27:44  &  7.40 & +0.20   &--0.40   &  2;4& --0.82   &$\pm$ 0.11 &         5       & 0.37 & 0.02\\ 
NGC339      &   00:57:45  &--74:28:21  &  9.80 & +0.08   &--0.10   &  7  & --1.50   &$\pm$ 0.14 &         7       & 0.18 & 0.02\\
NGC361      &   01:02:11  &--71:36:25  &  9.91 & +0.06   &--0.07   &  7  & --1.45   &$\pm$ 0.11 &         7       & 0.17 & 0.02\\ 
NGC411      &   01:07:56  &--71:46:09  &  9.15 & +0.06   &--0.07   &  1;3& --0.68   &$\pm$ 0.07 &         1;3     & 0.17 & 0.02\\ 
NGC416      &   01:07:58  &--72:21:25  &  9.84 & +0.06   &--0.08   &  7  & --1.44   &$\pm$ 0.12 &         7       & 0.20 & 0.02\\
\enddata
\tablenotetext{a}{Positions $\alpha_{2000}$ (hh:mm:ss), $\delta_{2000}$ (dd:mm:ss) are retrieved from 
Simbad Astronomical Database}
\tablenotetext{b}{The age for the clusters is given as log(age). The CMDs of the last three 
objects are providing only a crude age estimation, and the errors are not given.}
\tablenotetext{c}{The extinction information is retrieved from the website of the MCPS}
\tablerefs{1. \cite{as99}
2. \cite{chiosi95}
3. \cite{crowl01}
4. \cite{da costa98}
5. \cite{hill99}
6. \cite{hodge87}
7. \cite{mighell98}
8. \cite{mackey03b}
9. \cite{seggewiss89}}
\label{tab:smc}
\end{deluxetable}

\newpage
\begin{deluxetable}{lccrrrrcrrrr}
\tabletypesize{\small}
\tablenum{2}
\tablecolumns{12}
\tablecaption{LMC cluster sample.}
\tablehead{
\multicolumn{1}{l}{ID}&\colhead{$\alpha_{2000}$\tablenotemark{a}}&\colhead{$\delta_{2000}$\tablenotemark{a}}&
\multicolumn{4}{c}{Age, Errors \& Ref\tablenotemark{b}}&\multicolumn{3}{c}{[Fe/H], Err \& Ref}&
\colhead{$A_v$\tablenotemark{c}}&\colhead{$A_v$ err\tablenotemark{c}}}
\startdata 
NGC1466 & 03:44:33   &  --71:40:18  &  10.10 &+0.01 &--0.01  & 5  &  --2.17 &$\pm$0.20  &  12   &    0.39  &  0.02\\ 
NGC1651 & 04:37:32   &  --70:35:06  &  9.30  &+0.08 &--0.10  & 5  &  --0.37 &$\pm$0.20  &  12   &    0.35  &  0.05\\ 
NGC1711 & 04:50:37   &  --69:59:06  &  7.70  &+0.05 &--0.05  & 2  &  --0.57 &$\pm$0.17  &  2    &    0.56  &  0.01\\
NGC1718 & 04:52:25   &  --67:03:06  &  9.30  &+0.30 &--0.30  & 4  &  --0.42 &\nodata    &  20   &    0.51  &  0.06\\
NGC1754 & 04:54:17   &  --70:26:30  &  10.19 &+0.06 &--0.07  & 11 &  --1.54 &$\pm$0.20  &  12   &    0.40  &  0.04\\
NGC1777 & 04:55:48   &  --74:17:00  &  9.08  &+0.12 &--0.18  & 5  &  --0.35 &$\pm$0.20  &  12   &    0.39  &  0.02\\
NGC1786 & 04:59:06   &  --67:44:42  &  10.18 &+0.01 &--0.01  & 5  &  --1.87 &$\pm$0.20  &  12   &    0.62  &  0.04\\
NGC1805 & 05:02:21   &  --66:06:42  &  7.00  &+0.30 &--0.10  & 1  &  --0.25 &\nodata    &  1;9  &    0.32  &  0.02\\
\enddata
\tablenotetext{*}{The notes are the same as in Table~\ref{tab:smc}. The ages and metallicities for the last 11 objects are uncertain.}
\tablerefs{1. \cite{de grijs02} 2. \cite{dirsch00} 3. \cite{elson91} 4. \cite{elson88} 5. \cite{geisler97}
6. \cite{hill00} 7. \cite{hunter95} 8. \cite{jasniewicz94} 9. \cite{johnson01} 10. \cite{oliori98}
11. \cite{olsen98} 12. \cite{olszewski91} 13. \cite{rich01} 14. \cite{sirianni00} 15. \cite{suntzeff92}
16. \cite{seggewiss89} 17. \cite{bica98} 18. \cite{geisler97aj} 19. \cite{beasley02} 20.\cite{mackey03a}}
\label{tab:lmc}
\end{deluxetable}

\newpage
\begin{deluxetable}{lrrrr}
\rotate
\tabletypesize{\scriptsize}
\tablenum{3}
\tablecolumns{5}
\tablecaption{2MASS Atlas Images of SMC clusters.}
\tablehead{\multicolumn{1}{l}{Cluster
ID}&\colhead{N}&\colhead{$J$}&\colhead{$H$}&\colhead{$K_s$}}
\startdata 
NGC121 & 3 & s4\_aJ\_asky\_981020s0400256.fits & s5\_aH\_asky\_981020s0400256.fits & s6\_aK\_asky\_981020s0400256.fits\\
NGC152 & 3 & s7\_aJ\_asky\_980808s0580221.fits & s8\_aH\_asky\_980808s0580221.fits & s9\_aK\_asky\_980808s0580221.fits\\
NGC330 & 2 & s4\_aJ\_asky\_980809s0930245.fits & s5\_aH\_asky\_980809s0930245.fits & s6\_aK\_asky\_980809s0930245.fits\\
NGC339 & 3 & s7\_aJ\_asky\_980809s0940115.fits & s8\_aH\_asky\_980809s0940115.fits & s9\_aK\_asky\_980809s0940115.fits\\
NGC361 & 2 & s4\_aJ\_asky\_981021s0450256.fits & s5\_aH\_asky\_981021s0450256.fits & s6\_aK\_asky\_981021s0450256.fits\\
NGC411 & 3 & s7\_aJ\_asky\_981021s0560009.fits & s8\_aH\_asky\_981021s0560009.fits & s9\_aK\_asky\_981021s0560009.fits\\
NGC416 & 1 & s1\_aJ\_asky\_980809s1060021.fits & s2\_aH\_asky\_980809s1060021.fits & s3\_aK\_asky\_980809s1060021.fits\\
NGC419 & 3 & s1\_aJ\_asky\_980809s1060044.fits & s2\_aH\_asky\_980809s1060044.fits & s3\_aK\_asky\_980809s1060044.fits\\
NGC458 & 2 & s4\_aJ\_asky\_981021s0680021.fits & s5\_aH\_asky\_981021s0680021.fits & s6\_aK\_asky\_981021s0680021.fits\\
KRON3  & 2 & s4\_aJ\_asky\_980808s0470044.fits & s5\_aH\_asky\_980808s0470044.fits & s6\_aK\_asky\_980808s0470044.fits\\
LINDSAY1  & 3 & s1\_aJ\_asky\_980808s0210068.fits & s2\_aH\_asky\_980808s0210068.fits & s3\_aK\_asky\_980808s0210068.fits\\
          &  & s4\_aJ\_asky\_980808s0210080.fits & s5\_aH\_asky\_980808s0210080.fits & s6\_aK\_asky\_980808s0210080.fits\\
LINDSAY113& 4 & s7\_aJ\_asky\_981008s0590186.fits & s8\_aH\_asky\_981008s0590186.fits & s9\_aK\_asky\_981008s0590186.fits\\       
\enddata
\label{tab:smcAtl}
\end{deluxetable}

\newpage
\begin{deluxetable}{lrrrr}
\rotate
\tabletypesize{\scriptsize}
\tablenum{4}
\tablecolumns{5}
\tablecaption{2MASS Atlas Images of LMC clusters.}
\tablehead{\multicolumn{1}{l}{Cluster
ID}&\colhead{N}&\colhead{$J$}&\colhead{$H$}&\colhead{$K_s$}}
\startdata 
   NGC1466  &    2  &    s1\_aJ\_asky\_981027s0940021.fits   &    s2\_aH\_asky\_981027s0940021.fits     &   s3\_aK\_asky\_981027s0940021.fits\\
   NGC1651  &    2  &    s1\_aJ\_asky\_991026s1140068.fits   &    s2\_aH\_asky\_991026s1140068.fits     &   s3\_aK\_asky\_991026s1140068.fits\\    
   NGC1711  &    2  &     s1\_aJ\_asky\_980401s0140186.fits  &     s2\_aH\_asky\_980401s0140186.fits    &   s3\_aK\_asky\_980401s0140186.fits\\ 
   NGC1718  &    1  &     s1\_aJ\_asky\_981025s0810056.fits  &     s2\_aH\_asky\_981025s0810056.fits    &   s3\_aK\_asky\_981025s0810056.fits\\
   NGC1754  &    3  &     s1\_aJ\_asky\_981025s0820068.fits  &     s2\_aH\_asky\_981025s0820068.fits    &   s3\_aK\_asky\_981025s0820068.fits\\
   NGC1777  &    3  &     s1\_aJ\_asky\_981202s0710162.fits  &     s2\_aH\_asky\_981202s0710162.fits    &   s3\_aK\_asky\_981202s0710162.fits\\
   NGC1786  &    3  &     s1\_aJ\_asky\_981025s0860186.fits  &     s2\_aH\_asky\_981025s0860186.fits    &   s3\_aK\_asky\_981025s0860186.fits\\
   NGC1805  &    2  &     s1\_aJ\_asky\_981025s0950009.fits  &     s2\_aH\_asky\_981025s0950009.fits    &   s3\_aK\_asky\_981025s0950009.fits\\
\enddata
\label{tab:lmcAtl}
\end{deluxetable}

\newpage
\begin{deluxetable}{lrrrrrrrrrrr}
\tabletypesize{\small}
\tablenum{5}
\tablecolumns{8}
\tablecaption{Photometry of SMC clusters.}
\tablehead{\multicolumn{1}{l}{Cluster
ID}&\colhead{$\alpha_{2000}$}&\colhead{$\delta_{2000}$}&\colhead{$d$\arcsec}&\colhead{Flag}&\colhead{$R$\arcsec}&\colhead{$J$}
&\colhead{$Jerr$}&\colhead{$H$}&\colhead{$Herr$}&\colhead{$K$}&\colhead{$Kerr$}}
\startdata 
LINDSAY1   & 00:03:55.2 & -73:28:12.4 &  16.5  &  BBB   &  20  & 14.63  &  0.06  & 14.37  &  0.12 &  14.97  &  0.27\\                                
           &            &             &        &        &  40  & 12.60  &  0.03  & 12.23  &  0.04 &  12.27  &  0.05\\
           &            &             &        &        &  60  & 11.67  &  0.03  & 11.29  &  0.03 &  11.17  &  0.04\\
           &            &             &        &        &  80  & 10.88  &  0.02  & 10.42  &  0.03 &  10.30  &  0.03\\
           &            &             &        &        & 100  & 10.52  &  0.03  & 10.04  &  0.03 &   9.97  &  0.03\\
           &            &             &        &        & 120  & 10.35  &  0.03  &  9.87  &  0.03 &   9.82  &  0.03\\
           &            &             &        &        & 140  & 10.25  &  0.04  &  9.79  &  0.04 &   9.73  &  0.04\\
           &            &             &        &        & 160  & 10.09  &  0.04  &  9.61  &  0.04 &   9.57  &  0.04\\
\enddata
\tablenotetext{-}{Column~1 is 
the cluster designation, 2 and 3 are the right ascension and declination 
of the position used to center the apertures for the integral photometry. 
Column~4 is the offset of that position with respect of the cluster 
coordinates in SIMBAD, measured on 2MASS atlas images. Column~5 contains a flag, providing information about the age (first letter), metalicity estimates (second letter) and the photometry (third letter) for each cluster. A is corresponding to a reliable age, metallicity and photometry, B denotes the cases when the age and metallicity values are uncertain and when used in the third position B stands for the cases described in Section~\ref{notes} or if the photometry was provided in aperture sizes smaller than 200\arcsec. The radii of the
apertures used for each  
measurement are listed in column~6. The photometry information is given in 
columns 7 -- 12, in the order: $J$ magnitude, error of photometry in $J$, 
and the same information for the other two survey bands $H$ and $K_s$. The 
error of photometry in each band is calculated as a square root of the
quadrature sum of the zero point error, internal error of the photometry, 
and the error due to stochastic fluctuations of the background stellar
population.}
\label{tab:photsmc}
\end{deluxetable}

\newpage
\begin{deluxetable}{lrrrrrrrrrrr}
\tabletypesize{\small}
\tablenum{6}
\tablecolumns{8}
\tablecaption{Photometry of LMC clusters.}
\tablehead{\multicolumn{1}{l}{Cluster
ID}&\colhead{$\alpha_{2000}$}&\colhead{$\delta_{2000}$}&\colhead{$d$\arcsec}&\colhead{Flag}&\colhead{$R$\arcsec}&\colhead{$J$}
&\colhead{$Jerr$}&\colhead{$H$}&\colhead{$Herr$}&\colhead{$K$}&\colhead{$Kerr$}}
\startdata 
NGC1466  &  03:44:32.8  &  -71:40:16.6   &   2.5 &  AAA  &   20  & 10.94  &  0.01  & 10.44  &  0.01  & 10.35  &  0.01\\
         &              &                &       &       &   40  & 10.17  &  0.01  &  9.70  &  0.01  &  9.60  &  0.01\\
         &              &                &       &       &   60  &  9.83  &  0.02  &  9.39  &  0.01  &  9.30  &  0.01\\
         &              &                &       &       &   80  &  9.67  &  0.02  &  9.23  &  0.01  &  9.12  &  0.02\\
         &              &                &       &       &  100  &  9.58  &  0.03  &  9.15  &  0.02  &  9.03  &  0.02\\
         &              &                &       &       &  120  &  9.55  &  0.04  &  9.12  &  0.03  &  9.00  &  0.03\\
         &              &                &       &       &  140  &  9.52  &  0.05  &  9.08  &  0.03  &  8.95  &  0.04\\
         &              &                &       &       &  160  &  9.48  &  0.07  &  9.06  &  0.04  &  8.91  &  0.05\\
\enddata
\tablenotetext{-}{The notes are the same as in Table~\ref{tab:photsmc}}
\label{tab:photlmc}
\end{deluxetable}

\newpage
\begin{deluxetable}{lrrrrrrrrr}
\tabletypesize{}
\tablenum{7}
\tablecolumns{10}
\tablecaption{Comparison photometry of NGC152}
\tablehead{\colhead{Filter ID or color}&\colhead{$D\arcsec$}&\colhead{$X$}&\colhead{$Y$}&\colhead{offset}&\colhead{mag}&
\colhead{err}&\colhead{}&\multicolumn{2}{r}{\citeauthor{persson83}}}
\startdata
$K_s$     &   30 & 348   & 159   & 18.5  & 10.53  & $\pm$0.02 &  & 10.95 & $\pm$0.03 \\[3pt]
$J-K_s$  &        &          &          &          &  1.14   & $\pm$0.03 &  &  1.07  & $\pm$0.05 \\[3pt]
$H-K_s$ &        &          &          &          &  0.39   & $\pm$0.03 &  &  0.27  & $\pm$0.05 \\[6pt]
$K_s$    &  60  & 353   & 152   & 13.0  &  9.54 & $\pm$0.01 &  &  9.58 & $\pm$0.02 \\[3pt]
$J-K_s$ &        &          &          &          &  1.07 & $\pm$0.02 &  &  1.17 & $\pm$0.02 \\[3pt]
$H-K_s$&        &          &          &          &  0.38 & $\pm$0.01 &  &  0.37 & $\pm$0.03 \\
\enddata
\tablenotetext{-}{Column~1 is the Filter ID or color being compared with the results from \citeauthor{persson83}, the second column is showing the aperture used for that comparison, 3 and 4 are the atlas image coordinates of the recovered aperture centers based on the maximum throughput experiments. The offsets in pixels (or arc seconds) of these positions from the aperture center used in the present work are listed in column~5. Columns~6 and 7 are showing the magnitude or color values measured on the 2MASS images using recovered aperture positions from \cite{persson83} and their uncertainties. The corresponding values from \citeauthor{persson83} transformed into 2MASS magnitude system by the transformation equations of \cite{carpenter01} are listed in columns~8 and 9.§}
\label{tab:comp152}
\end{deluxetable}

\begin{deluxetable}{lrrrrrrrrr}
\tabletypesize{}
\tablenum{8}
\tablecolumns{10}
\tablecaption{Comparison photometry of NGC2209}
\tablehead{\colhead{Filter ID or color}&\colhead{$D\arcsec$}&\colhead{$X$}&\colhead{$Y$}&\colhead{offset}&\colhead{mag}&\colhead{err}
&\colhead{}&\multicolumn{2}{r}{\citeauthor{persson83}}}
\startdata
$K_s$   &   30 & 461   & 457   & 17.5  & 10.04 & $\pm$0.02 &  & 10.02    & $\pm$0.03 \\[3pt]
$J-K_s$ &       &       &       &       &  1.68 & $\pm$0.03 &  &  1.81    & $\pm$0.02 \\[3pt]
$H-K_s$ &       &       &       &       &  0.67 & $\pm$0.03 &  &  0.70    & $\pm$0.03 \\
\enddata
\tablenotetext{-}{The notes are the same as in Table~\ref{tab:comp152}}
\label{tab:comp2209}
\end{deluxetable}

\clearpage

\end{document}